\documentclass[aps,preprint]{revtex4-2}

\usepackage[english]{babel}

\usepackage{amsmath, amssymb}
\usepackage{graphicx}
\usepackage[colorlinks=true, allcolors=blue]{hyperref}

\def\be{\begin{equation}}
\def\ee{\end{equation}}
\def\bea{\begin{eqnarray}}
\def\eea{\end{eqnarray}}
\usepackage{xcolor}
\usepackage{physics}

\usepackage{braket}
\usepackage{dcolumn}
\usepackage{tabularx}
\usepackage{bm}

\usepackage{color}
\usepackage{xcolor}
\usepackage[normalem]{ulem}
\definecolor{color}{rgb}{0.11,0.45,0.02}

\renewcommand{\dd}{\operatorname{d\!}{}}

\begin{document}
\newcount\timehh  \newcount\timemm
\timehh=\time \divide\timehh by 60
\timemm=\time
\count255=\timehh\multiply\count255 by -60 \advance\timemm by \count255

\def\affiIOFFE{Ioffe Institute, Russian Academy of Sciences, 194021 St.~Petersburg, Russia}

\title{Non-perturbative macroscopic theory of interfaces with discontinuous dielectric constant }

\author{Y.~M.~Beltukov}
\email[]{yaroslav.beltukov@mail.ioffe.ru }
\author{A.~V.~Rodina}
\email[]{anna.rodina@mail.ioffe.ru}
\affiliation{Ioffe Institute, Russian Academy of Sciences, 194021 St.~Petersburg, Russia}
\author{ A.~Alekseev }
\email[]{Anton.Alekseev@unige.ch}
\affiliation{Department of Mathematics, University of Geneva, CH-1211 Genève 4, Switzerland}
\author{ Al.~L.~Efros }
\email[]{alex.l.efros.civ@us.navy.mil}
\affiliation{Naval Research Laboratory, Washington, DC 20375, USA}

\begin{abstract}
Discontinuity of dielectric constants at the interface is a common feature of all nanostructures and semiconductor heterostructures. 
Near such interfaces, a charged particle creates a singular self-interaction potential which may be attributed to interaction with fictitious mirror charges.  The singularity of this interaction at the interface presents an obstruction to a perturbative approach.
In several limiting cases, this problem can be avoided by zeroing out the carrier wave function at the interface.  In this paper, we have developed a non-perturbative theory which gives a self-consistent description of carrier propagation through an interface with a dielectric discontinuity. It is based on conservation of the current density propagating through the interface, and it is formulated in terms of general boundary conditions (GBC) for the wave function at the interface with a single phenomenological parameter $W$.  For these GBC, we find exact solutions  of the Schrödinger equation near the interface and the carrier energy spectrum including resonances.  Using these results, we describe  the photo effect at the semiconductor/vacuum interface and the energy spectrum of quantum wells (QWs)  at the interface with the vacuum or a high-k dielectric.  
For a surface of liquid helium, we estimate the parameter $W$, and match the resulting electron spectrum with the existing experimental data and theoretical analysis. 

\end{abstract}
\date{\today}
\maketitle

\section{Introduction}

The electronic, optical, and transport properties of nanostructures  with various shapes and their ensembles  are  mainly controlled by  spatial confinement and dielectric confinement effects.  The description of spatial confinements in spherical nanocrystals started in the early 1990s using the effective mass approximation \cite{EfrosEfros1982,Brus1983}, and now is a well-developed theoretical field.   The multiband effective mass theory can  take into account both a complex multiband structure and nonparabolicity of conduction and valence bands in III-V \cite{JOSA93,PRBEF-Ros,AdvanceMater,SercVahala}, 
II-VI \cite{Inuk-Wise} and perovskite  nanostructures \cite{Blaker2018}. It is applicable  not only in spherical nanocrystals, but also in nanowires and   nanorods \cite{ShabaevEfros},  and in  nanoplatelets \cite{SandrineNPT}. Effective mass calculations are supported by multiple  first principles calculations that include DFT   and empirical pseudo-potential  calculations \cite{EPP,DFT,PawelPRB2010, Galli,Teen} of structures containing more than one million atoms.

One of the important tools in the effective mass theory is general boundary conditions (GBC) (see \cite{Ando1982}) applied at interfaces between materials with different properties. In this approach, the interface is characterized by a transfer matrix $T$ which relates the values of the wave function and of its normal derivative on the two sides of the interface. The requirement for the probability current to be conserved across the interface is equivalent to the property of the Hamiltonian of the system being self-adjoint, and the evolution of the system being a unitary operator. These requirements force the transfer matrix $T$ to be a real matrix (up to an irrelevant total phase factor). A detailed analysis of GBC for abrupt boundaries is given in \cite{Zhu1983,Ando1989,Rodina2002,Rodina2003,Rodina2006}.

In contrast to successes of the theory of the spatial confinement, a self-consistent theory of  dielectric confinement is limited to several specific cases.  
The main stumbling  block  for a  quantitative theoretical description of this phenomenon is taking into account the interaction of the electron with the surrounding dielectric medium. It renormalizes the Coulomb interaction between any two charges, and for each electron it induces an effective mirror charge situated across the boundary between materials with different dielectric constants. The interaction of the electron with its mirror charge is referred to as self-interaction, and it corresponds to a singular potential near the interface. In the case of a flat semiconductor-dielectric interface of two materials with dielectric constants $\varepsilon_1$ and $\varepsilon_2$, the electric charge $e$ at distance $z$ from such the interface leads to its polarization which can be described by a mirror charge creating the potential \cite{LL}:
\begin{equation}      \label{eq:U_mirror}
U_{\rm self}(z) = 
\begin{cases}
\dfrac{q_1}{z} & z<0,~{\rm semiconductor} \\
\dfrac{q_2}{z} & z >0,~{\rm vacuum~or~dielectric}.
\end{cases}
\end{equation}
Here $q_1$ and $q_2$ are the Coulomb constants given by
$$
q_1=-\frac{e^2}{4 \varepsilon_1} \frac{\varepsilon_1 - \varepsilon_2}{\varepsilon_1 + \varepsilon_2}, \hskip 0.3cm
q_2=-\frac{e^2}{4 \varepsilon_2} \frac{\varepsilon_1 - \varepsilon_2}{\varepsilon_1 + \varepsilon_2}.
$$
We will mainly focus on the case of $\varepsilon_1 > \varepsilon_2$  which is typical for semiconductor-vacuum and semiconductor-dielectric interfaces. In that case, the Coulomb constants $q_1$ and $q_2$ are negative, and $|q_2| > |q_1|$. Consequently, the mirror force inside a semiconductor is repulsive  leading to dielectric confinement of carriers in the nanostructure.
We will also briefly touch upon the important for practical applications case of Si/high-k dielectric interfaces  \cite{GAAFET}, where  $\varepsilon_1 < \varepsilon_2$. 

 \begin{figure}[h]
  \includegraphics[scale=1]{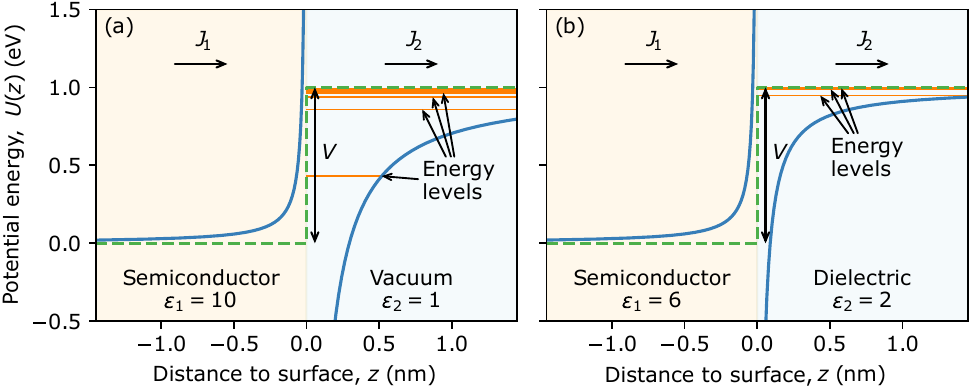}
\caption{The  heterostructure  formed  between  the semiconductor  and  the vacuum (a) or the dielectric matrix (b). 
The dashed line  shows  the step-like potential due to the band offset $V = 1$\,eV created by a semiconductor/vacuum (a) or by a semiconductor/dielectric matrix (b) interface.  Solid blue lines  show the sum of the step-like potential and the self-interacting potential, $U_{\rm self}(z)$, created  by mirror charge. The energy of the surface resonance levels (see Eq.~\eqref{eq:En})  created  by the attractive  potential $U_{\rm self}(z)$ in the vacuum or in the dielectric are shown  by horizontal orange lines.   
\label{fig:potential}}
\end{figure}

One simple model featuring  a discontinuity  of  the dielectric constant is the heterostructures shown in  Fig.~\ref{fig:potential}. In the absence of a dielectric discontinuity, it is represented  by a  $V$-height  step-like potential --- the band offset --- which is  shown by the dashed  lines. The  description of carrier scattering, reflection,  and propagation on such step-like potential is a standard textbook problem. At the same time, the mirror potential  $U_{\rm self}(z)$  described in Eq.\eqref{eq:U_mirror} gives rise to a $1/z$ divergence at $z=0$.

The diverging mirror potential $U_{\rm self}(z)$ cannot be consistently treated as a perturbation. Indeed, the first-order energy correction corresponding to  $U_{\rm self}(z)$ is given by the integral
$$
\Delta E=\int |\psi({\bm r})|^2 U_{\rm self}(z)\dd^3{\bm r},
$$
and this integral diverges at small $z$ unless one imposes the condition that the wave function $\psi({\bm r})$  vanishes at the interface
$\psi(S)=0$.  The problem exists for all shapes of nanostructures because   the mirror potential always has the same type of singularity at the nanostructure surface. 
In the multiband  approximation, the problem does not disappear even if one imposes the condition $\psi_i(S)=0$ on one of the components ({\em e.g.}\ the  $i$-component) of the wave functions.  In that case, the other components of the wave function will be  proportional to the derivative $\psi_i'(z) \neq 0$ leading to divergences in the integrals \cite{EranACSNano2022}. 

Uncertainty  in  the carrier wave function magnitude  at the interface  created by  $U_{\rm self}(z)$ does not allow to describe  its  leakage  into the matrix surrounding the nanostructures.  This leakage is crucial for description of transport in nanocrystal arrays and solids \cite{EriwnEfros,AntonErwinEfros}.  The absence of a coherent description of the leakage connected with mirror forces is even  more critical for the quantitative  description of the photo effect \cite{Alperovich2021,Alperovich2023}.

Problems related to the singularity $U_{\rm self} \sim 1/z$ have been addressed
in the literature. Note that the $1/z$ divergence in $U_{\rm self} \sim 1/z$ is unphysical \cite{LL} because one cannot use  the microscopic  potential  at  the inter-atomic distances to the interface.
To mimic  the {\it real} potential,  one can introduce  an intermediate region extending from $z=-d$ to $z=d$ (for some small $d$), and to have a smooth transitional potential in this region interpolating between the Coulomb like potentials on  the two sides of the interface. This procedure was first suggested in \cite{MuljarovSPIE1993,MuljarovPRB1995}, and later it was used to calculate the exciton binding energy in a quantum well  located near the semiconductor surface \cite{GipiusThihodeedv}.   
In a somewhat different approach, in order to avoid divergences in  the mirror potential in a dielectric quantum well, Kumagai and Takagahara \cite{TakagahraPRB2989} suggested to shift them by a small distance $\delta$ away from the interface, and they had shown that the dependence of the exciton binding energy on $\delta$ is very weak.  Yet another possible approach (see \cite{Frank}) is to introduce a  gradual change of the dielectric  constant  between the two  materials on some small inter atomic distance $d$.  Note however that all these approaches are {\em ad hoc}, and that their results depend on the particular model chosen to represent the interface.  

In this paper, we consider general boundary conditions (GBC) at the interface based on the conservation of the probability current across the interface: 
\begin{equation}
    J_{1,2}(z) = -i \frac{\hbar}{2m_{1,2}}\left( \psi_{1,2}^*\psi'_{1,2} - \psi_{1,2}\psi'^*_{1,2}\right).
\label{eq:3}
\end{equation}
Here $\psi_1$ and $\psi_2$ are the wave functions, $\psi'_1$ and $\psi'_2$ are their $z$-derivatives and $m_1$ and $m_2$ are the effective masses of electrons on the two sides of the interface. The divergence of the potential $U_{\rm self}(z)$ manifests  in the fact that the derivatives $\psi'_{1,2}(z)$ logarithmically diverge at the interface. Following \cite{Oliveira2009}, we introduce the modified derivatives $\tilde{\psi}'_{1,2}$ (see the next section for details) which are regular at the interface, and which can be used in the expression for the probability current.

Conservation of the probability current across the interface gives rise to a 3-parameter family of GBC described by the modified transfer matrix $\tilde{T}$. We will focus on the particular 1-parameter family suggested in \cite{Golovaty2019} which supplements the conservation of the probability current by the condition of continuity of the wave function at the interface.
These boundary conditions have the form
\begin{equation}        \label{eq:ourGBC}
\psi_2(0)=\psi_1(0), \hskip 0.3cm
    \frac{1}{m_2} \tilde{\psi}'_2(0)  = \frac{1}{m_1} \tilde{\psi}'_1(0) + \tilde{t}_{21} \psi_1(0) \, .
    \end{equation}
Here $\tilde{t}_{21}$ is the only nontrivial coefficient in the (modified) transfer matrix defining the GBC. It is convenient to parametrize it in terms of the parameter $W$ which has  dimensions energy$\cdot$length:
\begin{equation}     \label{eq:define W}
\tilde{t}_{21} = \frac{2W}{\hbar^2}.
\end{equation}
In very rough terms, $W$ can be thought of as the strength of an additional short-range potential $W\delta(z)$ at the interface.

We obtain an exact solution of the model presented  in Fig.~\ref{fig:potential}, and we apply the GBC \eqref{eq:ourGBC} to study electron scattering in the energy range $E>V$, resonances  and tunneling under the barrier for $V>E>0$, and the localized surface states for $E<0$. We observe a number of interesting new phenomena. For $E>V$, for some values of  the parameters the perfect transmission (zero reflection) can occur. Surprisingly, this effect may even occur at $E=V$ while without  a dielectric discontinuity the transmission coefficient vanishes at $E=V$.
In the region $V>E>0$ we encounter resonances in the reflection coefficient and tunneling of the wave function under the barrier near the resonance energies.
In the region $E<0$, the system may have discrete energy levels corresponding to quantum states localized near the interface.
 
As an application, we  study the effect of the parameter $W$  on  semiconductor photo-ionization, and  on electron confinement and wave function leakage  in a quantum well at the semiconductor surface. 
Again, we encounter several new phenomena. It turns out that the photo-ionization may be enhanced by taking into account the influence of the mirror potential $U_{\rm self}(z)$. For a quantum well, we observe a strong coupling between the quantum levels localized in the well and the surface states  created by the attractive mirror potential.

The structure of the paper is as follows:  in Section \ref{Sec.II}, using the conservation of the current density we  derive  the general  boundary conditions.   In Section \ref{sec:Coulomb},  we find exact solutions for the carrier wave functions and their asymptotics for $z\to 0$ and for $z \to \pm \infty$.   In Section \ref{sec:Eregs}, we study electron scattering at the interface in the energy range $E>V$, reflection and tunneling under the barrier for  $V>E>0$,  and surface localized states for $E<0$. We also introduce universal  equations  that describe scattering resonances and  localized states.   In Section \ref{Sec:5}, we apply our theory to  describe efficiency of  electron photoemission as well as the energy level structure in  surface quantum well, and their dependence on the surface parameter $W$. In Section \ref{Sec:discussion}, we outline possible future applications of our theory.


\section{Boundary conditions for interfaces with discontinuous dielectric constant}
\label{Sec.II}

In this Section, we consider Hamiltonians of the form
\be \label{eq:4}
\hat{H}=
\begin{cases}
    -\dfrac{\hbar^2}{2 m_{1} } \dfrac{\dd^2}{\dd z^2} +\dfrac{q_{1} }{z} +U^{\rm reg}_{1}(z), &
    {\rm for} \,\, z<0, \\[1em]
    -\dfrac{\hbar^2}{2 m_{2} } \dfrac{\dd^2}{\dd z^2} +\dfrac{q_{2} }{z} +U^{\rm reg}_{2}(z), &
    {\rm for} \,\, z>0.
\end{cases}
\ee
Here the region $z<0$ is a semiconductor, the region $z>0$ is the vacuum or a dielectric, and the interface between the two materials is located at $z=0$.
The potentials $U^{\rm reg}_{1,2}(z)$ are regular functions of $z$. Independently of the exact form of $U^{\rm reg}_{1,2}(z)$, eigenfunctions of Hamiltonians \eqref{eq:4} admit the following asymptotic behavior near $z=0$:
%
\begin{equation}
    \psi_{1,2}(z) = a_{1,2} + b_{1,2}z + c_{1,2} \,z \ln\left(\frac{|z|}{\lambda_{1,2}}\right) + d_{1,2} \, z^2 \ln\left(\frac{|z|}{\lambda_{1,2}}\right) +\dots .
\label {eq:5}
\end{equation}
Here we denote by $\psi_1(z)$ the wave function for $z<0$, and by $\psi_2(z)$ the wave function for $z>0$.
We have introduced characteristic length parameters 
$\lambda_{1,2}$ which enter the logarithms. One can see that the magnitudes of  $\lambda_{1,2}$ do not affect the logarithmic  divergence  of the wave functions  at $z\rightarrow  0$ in  Eq.~\eqref{eq:5}. We will discuss possible choices of $\lambda_{1,2}$ in the end of  this Section. 

If $\psi_{1,2}(0)=a_{1,2} \neq 0$, the terms proportional to $z \ln(|z|)$ and $z^2 \ln(|z|)$ in Eq.~\eqref{eq:5} are needed to ensure that $\hat{H}\psi(z)$ has no singularity  at  $z=0$. This condition implies the following relations on coefficients:
\begin{equation}    \label{eq:c_d}
c_{1,2} = 2 \, \frac{m_{1,2}q_{1,2}}{\hbar^2} a_{1,2}, \hskip 0.3cm d_{1,2} = 2 \left(\frac{m_{1,2}q_{1,2}}{\hbar^2}\right)^2 a_{1,2}.
\end{equation}
 Furthermore, using  Eq.~\eqref{eq:5}, we obtain the asymptotic behavior of derivatives of the wave function for $z<0$ and for $z>0$:
 \be
\psi'_{1,2}(z) = (b_{1,2}+c_{1,2}) + d_{1,2} \, z + c_{1,2} \, \ln\left(\frac{|z|}{\lambda_{1,2}}\right) + 2d_{1,2}\, z \, \ln\left(\frac{|z|}{\lambda_{1,2}}\right)+ \dots,
\label{eq:8A}
\ee
that is divergent at $z=0$ because the third  term in Eq. \eqref{eq:8A}  is proportional to $\ln(|z|)$. This is the reason to introduce the modified derivatives
\begin{equation}
    \tilde{\psi}'_{1,2}(z) = \psi'_{1,2}(z) - 2 \, \frac{m_{1,2}q_{1,2}}{\hbar^2} \psi_{1,2}(z) \, \ln\left(\frac{|z|}{\lambda_{1,2}}\right)
\label{eq:8}
\end{equation}
which have no singularities near $z=0$.  It is crucial that the probability current defined in Eq.~\eqref{eq:3}  can  be  expressed using the modified derivatives:
\be
J_{1,2}(z)\equiv -i{\hbar\over 2m_{1,2}}\Bigl(\psi_{1,2}^*(z) \tilde{\psi'}_{1,2}(z)  - \psi_{1,2} (z) \tilde{\psi}'^*_{1,2}(z) \Bigr)
\label{eq:9A}
\ee
The continuity of this probability current across the interface gives rise to a 3-parameter family of self-adjoint  GBCs:    
\begin{equation}      \label{eq:T_intro}
\left(
    \begin{array}{c}
    \psi_2 \\
   \tilde{\psi'}_{2}(z)/m_2
    \end{array}
    \right)= \tilde{T}_{21}
\left(
    \begin{array}{c}
    \psi_1 \\
   \tilde{\psi'}_{1}(z)/m_1
    \end{array}
    \right),~{\rm where}~ \det[\tilde{T}_{21}]=\tilde{t}_{11}\tilde{t}_{22}-\tilde{t}_{12}\tilde{t}_{21}=1.
\end{equation}
Here the matrix elements  $\tilde{t}_{ij}$ of the modified transfer matrix $\tilde{T}_{21}$ are characteristic of the interface region.
If the Coulomb constants $q_{1,2}$ vanish, the modified derivatives become ordinary derivatives, and the transfer matrix $\tilde{T}_{21}$ defines the usual GBC at the interface. 
In the presence of $q_{1,2}$, we would like to connect the matrix elements of the phenomenological  transfer matrix $\tilde{T}_{21}$ with  mirror force potential  without  necessarily going into microscopic calculations.

In order to address this question,  Golovaty \cite{Golovaty2019}  introduced an  interface layer of thickness $2d$ with the short-range  
potential
\begin{equation}     \label{eq:U_sing}
    U^{\rm sing}(z)=\frac{1}{d^2} \, U(z/d)
+ \frac{\ln\left(d/\lambda\right)}{d} \, Q(z/d) + \frac{1}{d} \, V(z/d).
\end{equation}
Here $\lambda$ is a characteristic length (which may be equal to $\lambda_1$ or $\lambda_2$).
 Regular functions $U$, $Q$ and $V$ are defined on the segment from $-1$ to $1$, and in Eq.~\eqref{eq:U_sing} they are scaled to the interface layer extending from $-d$ to $d$. These three terms in $U^{\rm sing}(z)$ give rise to three different singularities in the limit  $d \to 0$. In this limit, one of the following two things happens: either the domain splits into two non-interacting parts for $z<0$ and $z>0$ with Dirichlet boundary conditions ($\psi_1(0)=0$ and $\psi_2(0)=0$), or one obtains boundary conditions of the form \cite{Golovaty2019}
\begin{equation} \label{eq:Golobaty_bc_generalA}
\psi_2(0) = \theta \, \psi_1(0), \hskip 0.3cm
\frac{\theta}{m_2} \, \tilde{\psi}'_2(0) = \frac{1}{m_1} \tilde{\psi}'_1(0) + \frac{2W}{\hbar^2} \psi_1(0)   
\end{equation}
which correspond to the transfer matrix
\be
\tilde{T}_{21} =
\left(
\begin{array}{cc}
\theta & 0 \\
2W/\hbar^2\theta & 1/\theta
\end{array}
\right).
\ee

The mirror force potential in Eq.~\eqref{eq:U_mirror} grows as $z^{-1} \sim d^{-1}$ towards the interface layer. In comparison, the term $d^{-2}U( z/d) \sim d^{-2}$ grows much faster than the mirror force potential.
 Assuming that the potential within the boundary layer interpolates between the two mirror force potentials for $z<-d$ and for $z>d$ we consider it unlikely that the term $d^{-2}U( z/d) \sim d^{-2}$ arises in the limit of $d \to 0$. Hence, we will focus our attention on the simplified model with $U=0$.

 In that case, one obtains the Dirichlet boundary conditions $\psi_1(0)=\psi_2(0)=0$ unless the following condition is verified:
\begin{equation}     \label{eq:Golovaty_condition0}
\frac{1}{d} \int_{-d}^d Q(z/d) \dd z =
\int_{-1}^1 Q(y) \dd y = q_2-q_1.
\end{equation}
 If the condition \eqref{eq:Golovaty_condition0} is satisfied, in the limit $d \to 0$ one obtains the boundary conditions that preserve the continuity of the wave function at the interface:
\begin{equation}     \label{eq:Golovaty_bc_specialA}
\psi_2(0) =  \psi_1(0), \hskip 0.3cm
 \frac{1}{m_2} \, \tilde{\psi}'_2(0) = \frac{1}{m_1} \,\tilde{\psi}'_1(0) + \frac{2W}{\hbar^2}\, \psi_1(0). 
\end{equation}
Here $\theta =1$, and the parameter
\be
W = \frac{1}{d} \int_{-d}^d V( z/d) \dd z + q_1 \ln\left( \frac{\lambda}{\lambda_1} \right) - q_2 \ln\left( \frac{\lambda}{\lambda_2} \right)
\ee
combines the integral of the term $ d^{-1} V( z/d)$ across the interface layer with correction terms related to our choice of length parameters. In the limit $d \to 0$, it can be interpreted as the coefficient in front of the term $W \delta(z)$ in the short-range potential \cite{Golovaty2019}.
For matrix entries of the transfer matrix, we have  $\tilde{t}_{11}=\tilde{t}_{11}=1$, $\tilde{t}_{12}=0$, and  $\tilde{t}_{21} = 2 W /\hbar^2$. 

Finally, the term $\ln( d/\lambda)\,Q( z/d)/d$
also scales faster than the mirror charge  potentials for $d \to 0$. Therefore, it is also unlikely that such a term naturally arises in a realistic interface. Hence, if we considered the limit $d \to 0$, all interfaces with $q_1 \neq q_2$ would violate the condition \eqref{eq:Golovaty_condition0}, and would have been impenetrable. 

In practice, however, we consider small but finite values of $d$.  This allows us to model the interface potential using the term
\be
U^{\rm sing}(z) = \frac{1}{d} \, V_0(z/d)
\ee
that scales in the same way as the mirror force potential.  At finite $d$ we can present $U^{\rm sing}(z)$ as a sum of two terms, using
$$
\frac{1}{d}\, V_0(z/d) = 
\frac{\ln(d/\lambda)}{d} \, Q(z/d) + \frac{1}{d} \, V( z/d ),
$$
where the function $Q$ satisfies the condition \eqref{eq:Golovaty_condition0}. This yields the following relation between the integral of the potential across the interface and the parameter $W$ in the boundary conditions:
\begin{equation} \label{eq:W}
W_0 = \frac{1}{d} \int_{-d}^d V_0\left( \frac{z}{d}\right)  \dd z =  q_2 \ln\left( \frac{d}{\lambda_2} \right) - q_1 \ln\left( \frac{d}{\lambda_1} \right) + W.
\end{equation}
The parameter $W_0$ is the bare microscopic parameter of the interface, while the parameter $W$ is the macroscopic characteristic of the interface which may possibly be observed.  The relation between them is given by the mirror force renormalization \eqref{eq:W}. 

Equation \eqref{eq:W} and the boundary conditions \eqref{eq:Golovaty_bc_specialA} can be reproduced assuming the validity of the effective mass approximation in the transition layer and continuity of the wave function. The Hamiltonian takes the form
 \begin{equation}     \label{eq:H_layer}
    \hat{H}_{\rm layer}=
    \begin{cases}
      -\dfrac{\hbar^2}{2 m_{1} } \dfrac{\dd^2}{\dd z^2} +\dfrac{q_{1} }{z} +U^{\rm reg}_{1}(z), &
    {\rm for} \,\, z<-d, \\[1em]
    -\dfrac{\dd}{\dd z} \dfrac{\hbar^2}{2 m(z)} \dfrac{\dd}{\dd z}  +
    \dfrac{\ln\left(d/\lambda\right)}{d} \, Q\left( z/d \right) + \dfrac{1}{d} \, V\left( z/d \right) +U^{\rm reg}(z), & {\rm for} \,\,|z|<d, \\[1em]
    -\dfrac{\hbar^2}{2 m_{2} } \dfrac{\dd^2}{\dd z^2} +\dfrac{q_{2} }{z} +U^{\rm reg}_{2}(z), &
    {\rm for} \,\, z>d,  
    \end{cases}
 \end{equation}
where $m(z)$ interpolates between $m_1$ for $z<-d$ and $m_2$ for $z>d$, and the kinetic energy is symmetrized to ensure  that the Hamiltonian  is  self-adjoint \cite{Morrow1984}. Taking the integral $\int_{-d}^d 
\hat{H}_{\rm layer}(z)\psi(z) \dd z$  and neglecting  variation of the wave function across the interface region, we compute
\be 
\frac{1}{m_2} \, \psi'_2(d) - \frac{1}{m_1} \, \psi'_1(-d)  = 
 \frac{2}{\hbar^2}\int_{-d}^{d} \left( \frac{1}{d} V_0\left(\frac{z}{d}\right) + U^{\rm reg}(z)-E\right) \psi(z) \dd z 
 =  \frac{2}{\hbar^2} \, W_0 \psi(0).
\label {eq:20}
\ee
Here we have neglected the term 
$$
\int_{-d}^{d} \left( U^{\rm reg}(z)-E\right) \dd z =
\int_{-d}^{d}  U^{\rm reg}(z) \dd z - 2Ed
$$
which is small in comparison to $W_0$ for small values of $d$.   Replacing the wave function derivatives  in Eq.~\eqref{eq:20} by the modified derivatives from Eq.~\eqref{eq:8}  we finally obtain
$$
\frac{1}{m_2} \, \tilde{\psi}'_2(d) - \frac{1}{m_1} \, \tilde{\psi}'_1(-d) 
 =  \frac{2}{\hbar^2}\left[W_0 + q_1\ln\left(\frac{d}{\lambda_1 }\right) - q_2\ln\left(\frac{d}{\lambda_2 }\right)\right]\psi(0) 
 =  \frac{2}{\hbar^2} W \psi(0).
$$
Due   to  convergence of the   modified derivatives $ \tilde{\psi}'_{1,2}(d)\rightarrow  \tilde{\psi}'_{1,2}(0)$  when $d\rightarrow 0$ we arrive at the GBC defined in Eq.~\eqref{eq:Golobaty_bc_generalA}.

Eq.~\eqref{eq:W} also allows us to estimate  the effect of dielectric constant differences   on the surface parameter $W$.   In more detail, the transition layer thickness  $2d$  should be on the order of inter-atomic distances.  The length  scale parameters  $\lambda_{1,2}$ in  $\ln(d/\lambda_{1,2})$  that control the electron behavior outside of the interface layer should  naturally satisfy the condition $d < \lambda_{1,2}$. One natural choice for characteristic lengths is the radii $\lambda_{1,2}=\hbar^2/(m_{1,2}|q_{1,2}|)$ of electron localization in the image potentials. 
Note that the characteristic lengths $\lambda_{1,2}$ are defined up to numerical factors $\lambda_{1,2} \to \nu_{1,2} \lambda_{1,2}$. Such rescalings result in the change of definitions of $\tilde{\psi}'_{1, 2}$  (see Eq.~\eqref{eq:8}) and of the parameter $W$ (see Eq.~\eqref{eq:W}), while all physical quantities remain the same. These redefinitions can be interpreted as finite renormalizations in the framework of renormalization theory.


It is important that  we apply boundary conditions to the effective  mass wave functions at the interface ($z=0$)  where they are well defined.  This allows us to use these wave functions for  calculations of various physical characteristics  within the effective mass approximation  without any restrictions.  At the same time, we do not know the microscopic details of the potential in the interface region and consider the surface parameter $W$ as the phenomenological one. 
\newline\newline

\section{ Exact solutions \label{sec:Coulomb}}

The Schrödinger equation for the Hamiltonian \eqref{eq:4} admits exact solutions on both sides of the interface. For an arbitrary complex energy $E$, they are given by
\begin{equation}      \label{eq:hyper1A}
\begin{array}{ll}
\psi_1(z) = z e^{\varkappa_1 z}\Bigl[C_1 M(1+\alpha_1, 2, -2\varkappa_1 z)  + D_1 U(1+\alpha_1, 2, -2\varkappa_1 z)\Bigr] & {\rm for} \, z<0 \\
\psi_2(z) = z e^{- \varkappa_2 z}\Bigl[C_2 M(1+\alpha_2, 2, 2\varkappa_2 z) + D_2 U(1+\alpha_2, 2, 2\varkappa_2 z)\Bigr] & {\rm for} \, z>0,
\end{array}
\end{equation}
where $M(a,b,z) = {}_1F_1(a,b,z)$  and $U(a,b,c)$ are the confluent hypergeometric   functions of the first (Kummer's) and of the second (Tricomi's) kind, respectively.  In Eq.~\eqref{eq:hyper1A}   $ \varkappa_2 = \sqrt{2m_2(V- E)}/\hbar$ for the energies  $E<V$  and  $\varkappa_1 = \sqrt{-2m_1 E}/\hbar$   for negative energies $E<0$.  For $E>0$ we have $\varkappa_1 = - ik_1$ with $k_1$ positive. Similarly, for $E>V$ we define $\varkappa_2=-ik_2$ with $k_2$ positive.
In addition, it is convenient to introduce the following notation:
\begin{equation}
\alpha_1 = q_1 \varkappa_1/2E, \hskip 0.3cm
\alpha_2= q_2 \varkappa_2/2(V - E).
\end{equation}
For $E>0$ we denote $\alpha_1 = - i \beta_1$, and for 
$E>V$ we denote $\alpha_2 = -i\beta_2$ with $\beta_1$ and $\beta_2$ positive.
 The wave functions \eqref{eq:hyper1A} should be accompanied by some boundary conditions at the interface. 

The asymptotic of   the wave functions \eqref{eq:hyper1A} for $z \to \pm \infty$ are given by:
\begin{align}
    \psi_1(z) &\simeq \frac{-1}{2\varkappa_1}\left(\frac{C_1 (-2\varkappa_1 z)^{\alpha_1}e^{-\varkappa_1z}}{\Gamma(1+\alpha_1)} - \frac{C_1 e^{\varkappa_1 x}}{(2\varkappa_1 z)^{\alpha_1}\Gamma(1-\alpha_1)} + \frac{D_1 e^{\varkappa_1 z}}{(-2\varkappa_1z)^{\alpha_1}}\right), \quad z\to-\infty, \nonumber\\
    \psi_2(z) &\simeq \frac{1}{2\varkappa_2}\left(\frac{C_2 (2\varkappa_2 z)^{\alpha_2}e^{\varkappa_2z}}{\Gamma(1+\alpha_2)} - \frac{C_2 e^{-\varkappa_2 z}}{(-2\varkappa_2 z)^{\alpha_2}\Gamma(1-\alpha_2)} + \frac{D_2 e^{-\varkappa_2 z}}{(2\varkappa_2z)^{\alpha_2}}\right), \quad z\to+\infty.
\label{eq:17}
\end{align}
For $z \to 0$, the  wave functions \eqref{eq:hyper1A} satisfy conditions \eqref{eq:5}, and they can be written as follows:
\begin{align}
    \psi_1(z) &\simeq C_1 z+ \frac{D_1}{\Gamma(\alpha_1)}\left(
    \frac{-1}{2\varkappa_1\alpha_1} + z \Bigl[
    \ln(-2\varkappa_1 z) - 1 + \sigma(\alpha_1)
    \Bigr]
    \right), \quad z\to-0, \nonumber\\
    \psi_2(z) &\simeq C_2 z + \frac{D_2}{\Gamma(\alpha_2)}\left(
    \frac{1}{2\varkappa_2\alpha_2} + z \Bigl[
    \ln(2\varkappa_2 z) - 1 + \sigma(\alpha_2)
    \Bigr]
    \right), \quad z\to+0.
\label{eq:18}
\end{align}
Here we introduce the function
\begin{equation}
    \sigma(\alpha) = 2\gamma + \frac{\Gamma'(\alpha)}{\Gamma(\alpha)} + \frac{1}{2\alpha},
\end{equation}
where $\gamma\approx 0.577216$  is  Euler's constant. The function $\sigma(\alpha)$ has the following properties:
\begin{equation}
    \sigma(\alpha) - \sigma(-\alpha) = -\pi \cot(\pi \alpha), \hskip 0.3cm
    \sigma(i\beta) - \sigma(-i\beta) = i\pi \coth(\pi \beta).  \label{eq:sigma_prop}
\end{equation}
It has  simple poles at non-positive integers:
\begin{align}
    \sigma(\alpha) &\simeq -\frac{1}{2\alpha} + \gamma
    \,\, {\rm for} \,\, \alpha \to 0, \nonumber\\
    \sigma(\alpha) &\simeq -\frac{1}{\alpha + n} + \sigma(n)
    \,\, {\rm for} \,\, \alpha \to -n,
\end{align}
where $n$ is a positive integer and $\sigma(n) = \gamma - 1/(2n) +\sum_{k=1}^n 1/k$  has a finite value.

\begin{figure}
    \centering
    \includegraphics[scale=1]{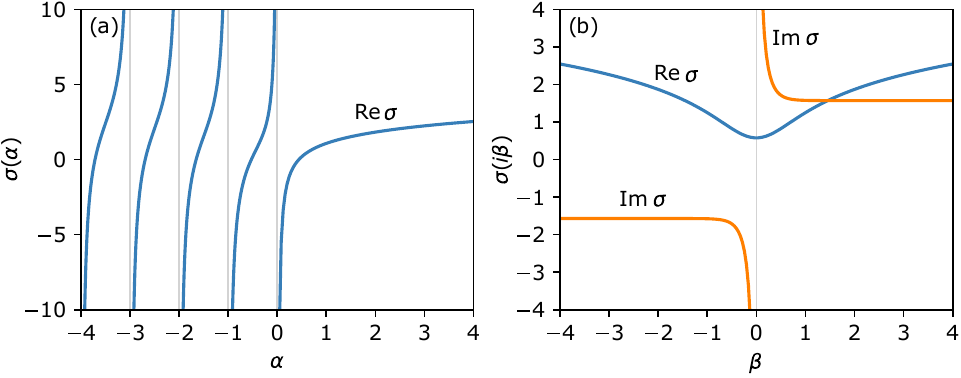}
    \caption{Function $\sigma(x)$ for (a) real ($x=\alpha$) and  (b) imaginary ($x=i\beta$) arguments. Vertical gray lines show poles of $\sigma(x)$.}
    \label{fig:small_sigma}
\end{figure}

The imaginary and real parts of 
    $\sigma(\alpha)$  are shown in Fig.~\ref{fig:small_sigma} as a function of $\alpha$. 

 
Using Eqs.~\eqref{eq:8} and \eqref{eq:18} one can calculate the modified derivatives at $z=0$:
\begin{equation}
\begin{array}{lll}
    \tilde{\psi}_1'(0)  & = & C_1 + \frac{D_1}{\Gamma(\alpha_1)}\Bigl(
    \ln(2\varkappa_1 \lambda_1) + \sigma(\alpha_1) \Bigr), \\
    \tilde{\psi}_2'(0)  & =  &C_2 + \frac{D_2}{\Gamma(\alpha_2)}\Bigl(
    \ln(2\varkappa_2 \lambda_2) + \sigma(\alpha_2) \Bigr).
    \end{array}
    \label{eq:psi1der_tilde1}
\end{equation}
  Using boundary conditions defined in Eq.~\eqref{eq:Golovaty_bc_specialA},  we  can find the energy spectra and the corresponding  wave functions.

Let us start from the limit $W \to \pm \infty$ that we have briefly discussed in the previous section.  The problem splits into two non-interacting parts corresponding to Dirichlet boundary  condition: $\psi_1(0)=0$ and  $\psi_2(0)=0$ and may give rise to isolated eigenvalues of the problem. These boundary conditions imply $J(0)=0$, and there is no probability current through the interface $z=0$.
Using the asymptotic   behavior described by Eq.~\eqref{eq:18} we obtain
\begin{equation}
    \psi_1(0) = -\frac{D_1}{2\varkappa_1\alpha_1 \Gamma(\alpha_1)}, \hskip 0.3cm
    \psi_2(0) = \frac{D_2}{2\varkappa_2\alpha_2  \Gamma(\alpha_2)}.
\end{equation}
 If $q_1, q_2 <0$ the mirror force  for $z>0$  is attractive and leads to isolated eigenstates.  The energy levels  are connected with the poles of the $\Gamma$-function $\alpha_2=-n$. The energy levels  are  given by:
\begin{equation}   \label{eq:En} 
    E_n=V - \frac{\mathcal{R}_2}{n^2}, \quad \mathcal{R}_2 = \frac{m_2q_2^2}{2\hbar^2} = \frac{\hbar^2}{2 m_2 \lambda_2^2}.
\end{equation}
We will refer to the energies $E_n$  as the Dirichlet spectrum; the corresponding levels are shown in Fig.~\ref{fig:potential}. 

\begin{figure}
    \centering
    \includegraphics[scale=1]{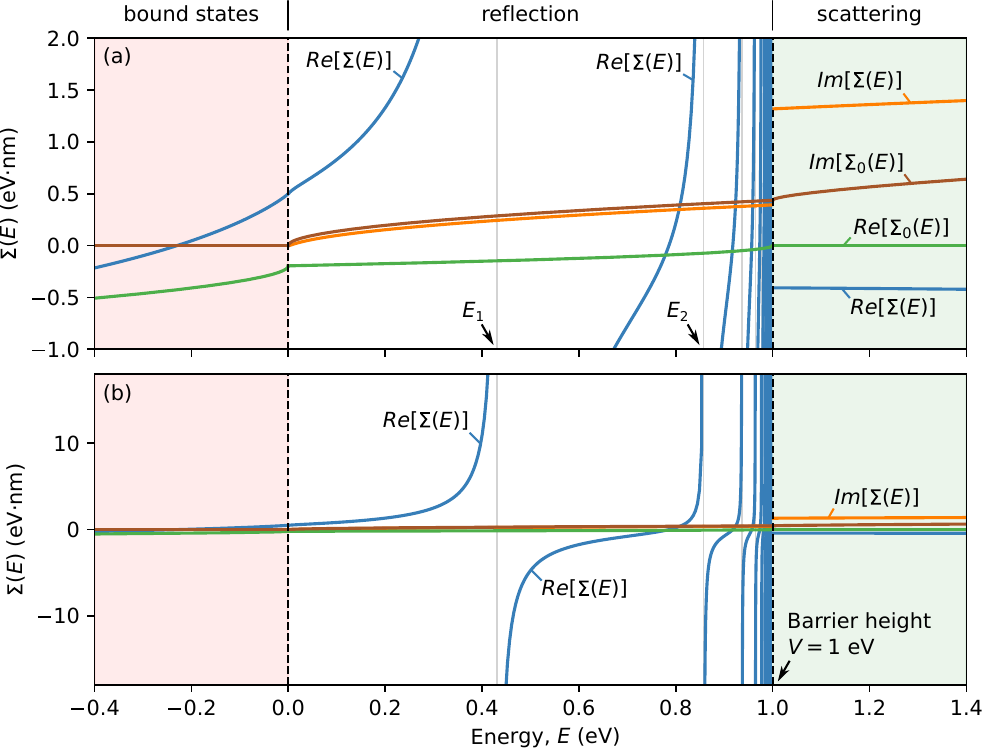}
    \caption{Dependence of $\Sigma(E)$ and $\Sigma_0(E)$ on the energy $E$, calculated for barrier height $V=1$\, eV. Vertical gray lines show poles $E = E_n$. Panels (a) and (b)  differ in vertical scale only. The range $E<0$ where the surface bound  states might exist is marked by the  rosa color; in the white region $0<E<V$ the under barrier reflection takes place; in the green region $E>V$ the scattering and reflection above the barrier occurs.  }
    \label{fig:Sigma}
\end{figure}

  Now let us consider the solution at  finite  ${W} $.   Imposing the boundary conditions described in Eq.~\eqref{eq:Golovaty_bc_specialA}  we obtain the relation between the coefficients $C_{1,2}$ and $D_{1,2}$:
\begin{equation}
    D_1 = \left(\frac{C_1}{m_1} - \frac{C_2}{m_2}\right) \frac{m_1q_1 \Gamma(\alpha_1)}{\Sigma(E) - {W}}, \quad   
    D_2 = \left(\frac{C_1}{m_1} - \frac{C_2}{m_2}\right) \frac{m_2q_2 \Gamma(\alpha_2)}{\Sigma(E) - {W}}, \label{eq:D12a}
\end{equation}
where the complex function
\begin{equation}   \label{eq:Sigma_mainA}
    \Sigma(E) = 
    - q_1\bigl(\ln(2\varkappa_1 \lambda_1) + \sigma(\alpha_1)\bigr)
    + q_2\bigl(\ln(2\varkappa_2 \lambda_2) + \sigma(\alpha_2)\bigr)
\end{equation}
is shown in Fig.~\ref{fig:Sigma}.  Function    $\Sigma(E)$  plays a crucial role in the description of the  energy spectra at the interface.
In the energy region $E<0$,  Eq.~\eqref{eq:Sigma_mainA}   can be used directly since $\varkappa_{1,2}$ are real positive and $\alpha_{1,2}$ are real.  In the energy region $V>E>0$, the expressions for $\Sigma(E)$ Eq.~\eqref{eq:Sigma_mainA}  is valid with  $\varkappa_1 = -i k_1$ and $\alpha_1 = -i\beta_1$ and   can be rewritten as
\begin{equation}
    \Sigma(E) = -q_1\Bigl(\ln(2k_1 \lambda_1) - \frac{i\pi}{2} + \sigma(-i\beta_1)\Bigr)+ q_2\bigl(\ln(2\varkappa_2 \lambda_2) + \sigma(\alpha_2)\bigr).
\end{equation}
Finally,  in the energy region $E>V$  we can rewrite  $\Sigma(E)$ as
\begin{equation}
    \Sigma(E) = -q_1\Bigl(\ln(2k_1 \lambda_1) - \frac{i\pi}{2} + \sigma(-i\beta_1)\Bigr) + q_2\Bigl(\ln(2k_2 \lambda_2) - \frac{i\pi}{2} + \sigma(-i\beta_2)\Bigr).
\end{equation}
The imaginary part of $\Sigma(E)$ can be simplified using the identity $\Im \sigma(-i\beta) = -(\pi/2)\coth(\pi \beta)$ for real $\beta$ (which results from Eq.~\eqref{eq:sigma_prop}). This yields:
\begin{equation}
    \Im \, \Sigma(E) = \begin{cases}
        G_1(E) + G_2(E), & E > V, \\
        G_1(E), & 0 < E < V, \\
        0, & E < 0,
    \end{cases}
\end{equation}
where
\begin{align} \label{ImSigma}
    G_1(E) &= \frac{\pi q_1}{1-\exp(-2\pi\beta_1)} = \frac{\pi q_1}{1 - \exp(-\pi q_1 \sqrt{2m_1/E\hbar^2})}, \\
    G_2(E) &= \frac{-\pi q_2}{1-\exp(2\pi\beta_2)} =  \frac{- \pi q_2}{1 - \exp(\pi q_2 \sqrt{2m_2/(E - V)\hbar^2})}.
\end{align}
When  the dielectric constants are equal to each other and $q_1, q_2 \to 0$, the function $\Sigma(E)$  tends to the function $\Sigma_0(E)$ given by
\begin{equation}\label{eq:sigma0}
    \Sigma_0(E) = \begin{cases}
        i\hbar \sqrt{\frac{E}{2m_1}} + i\hbar \sqrt{\frac{E - V}{2m_2}}, 
        & E > V, \\
       i\hbar \sqrt{\frac{E}{2m_1}} - \hbar \sqrt{\frac{V - E}{2m_2}}, & 0 < E < V, \\
        - \hbar \sqrt{\frac{-E}{2m_1}} - \hbar \sqrt{\frac{V - E}{2m_2}}, & E < 0.
    \end{cases}
\end{equation}
The functions $\Sigma(E)$  and $\Sigma_0(E)$ are shown in Fig.~\ref{fig:Sigma}.  One can see that  $\Sigma(E)$ tends to $\Sigma_0(E)$ at high energy.

Note, that both $G_1(E)$ and $G_2(E)$  are positive because $q_1$ and $\beta_1$ have the same sign, while $q_2$ and $\beta_2$ have opposite signs,  and  $G_1(0) = \pi \max(q_1,0)$, $G_2(V) = \pi \max(-q_2,0)$. The finite value of $G_2(V)=-\pi q_2$ in the case $q_1,q_2<0$ is responsible for the jump of the ${\rm Im}\, \Sigma(E)$ at $E=V$ while there is no jump at $E=0$ (see in Fig.~\ref{fig:Sigma}). This jump is vanishing with $q_2 \to 0$, and it is absent in ${\rm Im}\, \Sigma_0(E)$.

\section{
Scattering, resonances, and discrete spectrum \label{sec:Eregs}}

In this Section, we consider  scattering, reflection, and propagation  of electrons through interfaces separating two solids with different dielectric constants  ($q_1\neq 0$ and $q_2\neq 0$)  as shown in Fig.~\ref{fig:potential}.  
We have a separate discussion for each of the  three energy regions: $E>V$, $V > E >0$ and $E<0$.

\subsection{ Reflection and propagation  for $E>V$}

We first consider the scattering problem for $E>V>0$. The asymptotics  of the wave functions for  $z \to \pm \infty$  can be written as:
\begin{equation}       \label{eq:planewaves_Coulomb}
\begin{array}{ll}
    \psi_1(z) & \simeq A_1 e^{i k_1 z - i\beta_1 \ln(-2k_1 z)} + B_1 e^{-i k_1 z+ i\beta_1 \ln(-2k_1 z)}, \\
    \psi_2(z) & \simeq A_2 e^{i k_2 z + i\beta_2 \ln(2k_2 z)} + B_2 e^{-i k_2 z - i\beta_2 \ln(2k_2 z)}.
    \end{array}
\end{equation}
Note  that  the plane wave asymptotic behavior in  Eq.~\eqref{eq:planewaves_Coulomb} has long-range logarithmic corrections from the Coulomb potential. The coefficients $A_{1,2}, B_{1,2}$ in Eq.~\eqref{eq:planewaves_Coulomb}  are related to  coefficients $C_{1,2}, D_{1,2}$ from
Eq.~\eqref{eq:17}  as follows:
\bea
    A_1 &=& -\frac{i C_1}{2k_1} \frac{e^{-\pi\beta_1/2}}{\Gamma(1-i\beta_1)}, ~~~~~~
    B_1 =\frac{i C_1}{2k_1}\frac{e^{-\pi\beta_1/2}}{\Gamma(1+i\beta_1)} - \frac{iD_1}{2k_1}e^{\pi\beta_1/2},  \nonumber \\
    A_2 &=& -\frac{iC_2}{2k_2}\frac{e^{-\pi\beta_2/2}}{\Gamma(1+i\beta_2)} + \frac{iD_2}{2k_2}e^{\pi\beta_2/2}, ~~~~
    B_2 = \frac{i C_2}{2k_2} \frac{e^{-\pi\beta_2/2}}{\Gamma(1-i\beta_2)}.
\label{eq:36}
\eea
Using Eq.~\eqref{eq:D12a}, we can express  four coefficients $A_{1,2}$, $B_{1,2}$ in terms  $C_{1,2}$. For arbitrary $C_{1,2}$ the 
relationships between  $A_{1,2}$, $B_{1,2}$ can be written with the help of the scattering matrix $S({W}, E)$:
\begin{equation}      \label{eq:scattering_Coulomb}
\left(
    \begin{array}{c}
  B_1 \\
   A_2
    \end{array}
    \right)= S({W}, E)
\left(
       \begin{array}{c}
  A_1 \\
   B_2
    \end{array}
    \right),
\end{equation}
where
\begin{equation}
    S({W}, E) = 
    \left(
    \begin{array}{cc}
    \frac{\Gamma(-i\beta_1)}{\Gamma(i\beta_1)} - \frac{iq_1\beta_1\Gamma^2(-i\beta_1)}{\Sigma(E) - {W}}e^{\pi\beta_1}
    &
    \frac{i q_2 \beta_1 \Gamma(-i\beta_1)\Gamma(-i\beta_2)}{\Sigma(E) - {W}} e^{\frac{\pi}{2}(\beta_1 + \beta_2)}
    \\
    -\frac{i q_1 \beta_2 \Gamma(-i\beta_1)\Gamma(-i\beta_2)}{\Sigma(E) - {W}} e^{\frac{\pi}{2}(\beta_1 + \beta_2)}
    &
    \frac{\Gamma(-i\beta_2)}{\Gamma(i\beta_2)} + \frac{iq_2\beta_2\Gamma^2(-i\beta_2)}{\Sigma(E) - {W}}e^{\pi\beta_2}
    \end{array}
    \right). \label {eq:38}
\end{equation}

For $q_{1,2} \to 0$, the scattering matrix  $S(W,E)$ tends to the limit $S_0(W, E)$ which describes scattering on the step-like barrier (without Coulomb potential): 
\begin{eqnarray}
S_0(W, E) = \frac{1}{\Sigma_0(E)-W} \
\left(  \begin{array}{cc}
i \frac{\hbar^2}{2}\left(\frac{k_1}{m_1}  - \frac{k_2}{m_2}\right) +W & i \hbar^2\frac{k_2}{m_2} \\ i\hbar^2 \frac{k_1}{m_1}  & i \frac{\hbar^2}{2}\left(\frac{k_2}{m_2} - \frac{k_1}{m_1}\right)+W 
    \end{array}
    \right) 
\end{eqnarray}

For $E>V$, the scattering matrix $S(W,E)$ satisfies the unitary condition:
\begin{equation}      \label{eq:39}
S({W}, E)^* \, K \, S({W}, E) = K, \hskip 0.3cm
K=
\left(
\begin{array}{cc}
k_1/m_1 & 0 \\
0 & k_2/m_2
\end{array}
\right).
\end{equation}
This condition follows from the conservation of the probability current on the two sides of the interface. 

The reflection coefficient $R$ and the transmission coefficient $T_{\rm tr}=1-R$ (defined as electron fluxes normalized to the incident flux \cite{Alperovich2021}) can be  expressed  via the elements of the scattering matrix: 
\begin{equation}
R({W}, E)=| S_{11}({W}, E)|^2=| S_{22}({W}, E)|^2 \, , \quad T_{\rm tr}({W}, E)=| S_{12}({W}, E)  S_{21}({W}, E)| \, .
\label{eq:40}
\end{equation}
Substituting the expressions for these matrix elements from  Eq.~\eqref{eq:38}  we obtain  the reflection coefficient
\begin{equation}
R({W}, E)  = 1 - \frac{4G_1(E)G_2(E)}{|\Sigma(E) - {W}|^2} = \frac{\bigl(\Re \Sigma(E) - {W}\bigr)^2 + \bigl(G_1(E) - G_2(E)\bigr)^2}{\bigl(\Re \Sigma(E) - {W}\bigr)^2 + \bigl(G_1(E) + G_2(E)\bigr)^2},
\label{eq:41}
\end{equation}
 and the transmission coefficient
\be\label{eq:T}
    T_{\rm tr}({W}, E) = 1 - R({W}, E) = \frac{4G_1(E)G_2(E)}{|\Sigma(E) - {W}|^2}.
\ee
\begin{figure}
    \centering
    \includegraphics[scale=1]{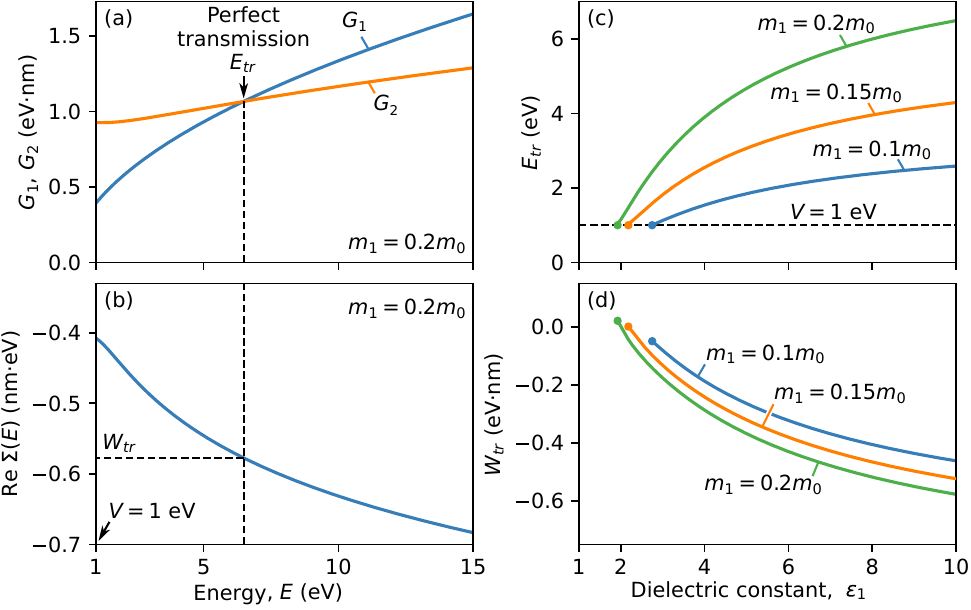}
    \caption{(a,\,b). 
    Calculation  of the perfect transmission energy $E_{\rm tr}>V$ and surface parameter $W_{\rm tr}$ in structures with dielectric constants $\varepsilon_2=1$ and $\varepsilon_1=10$,  and electron effective masses
 $m_2=m_0$, $m_1=0.2 m_0$. (c,\,d) Dependencies of $E_{\rm tr}$ and $W_{\rm tr}$ on $\varepsilon_1$ for $\varepsilon_2=1$, $m_2=m_0$ and different values of $m_1$. All calculations are  conducted for the barrier height  $V=1$~eV.}  
    \label{fig:G2}
\end{figure}%
\begin{figure}[h!]
    \centering
    \includegraphics[scale=1]{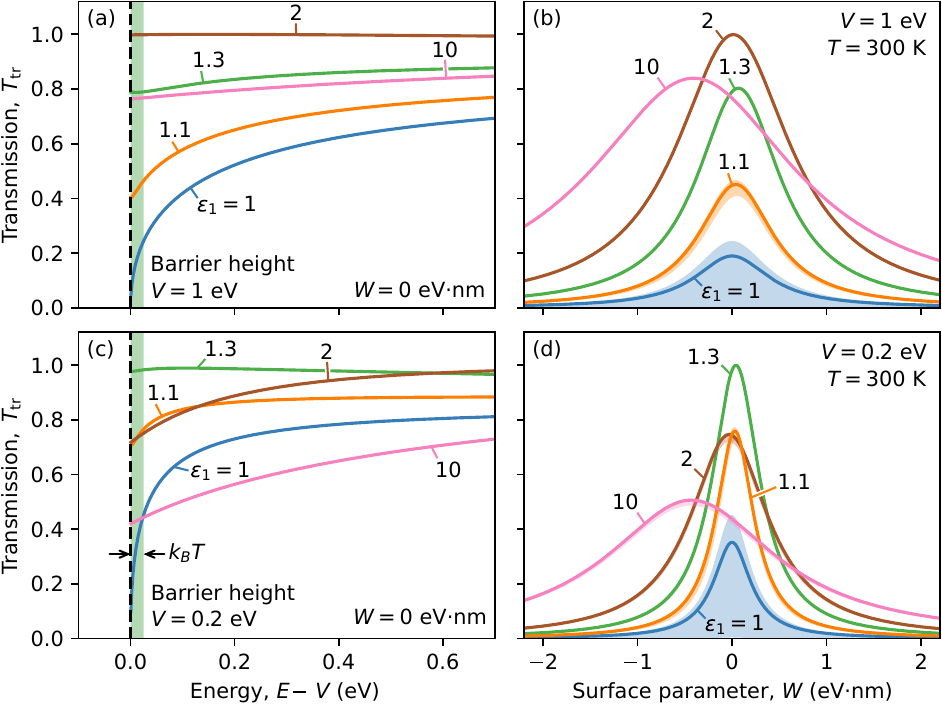}
  \caption{
  (a,\,c) Dependence of the transmission coefficient $T_{\rm tr}$ on energy $E-V$ for the surface parameter $W=0$. (b,\,d) Dependence of the averaged transmission efficiency $T^{\rm eff}_{\rm tr}$ on  the surface parameter $W$.  In (a,\,c) the  filled area  shows the energy interval  $V<E<V+k_{\rm B}T$ with $T=300$~K.   The fill up the area  in the  (b) and (d) panels   shows the room temperature distribution of  the transmission coefficient  $T_{\rm tr}$ in the same energy interval. The calculations are conducted for barrier height $V=1$~eV (a,\,b) and $V=0.2$~eV (c,\,d) and several dielectric constants  $\varepsilon_1$ and $\varepsilon_2=1$. }
    \label{fig:T} 
\end{figure}%
For complex values of $E$,  the scattering matrix $S(W,E)$ has poles corresponding to resonances at solutions of the equation $\Sigma(E)={W}$.
However, for ${\rm Re} \, E > V$ the imaginary part $\Im \, \Sigma(E)=G_1(E)+G_2(E)$ is always positive and does not  vanish even for complex energies $E$. One can observe maxima of the transmission coefficient (and minima of the reflection coefficient) near solutions of the equation  ${\rm Re} \, \Sigma(E)={W}$.   

In what follows, we discuss an interesting phenomenon of perfect transmission ($T_{\rm tr}=1$) which may occur due to dielectric confinement.

Perfect transmission (that is, zero reflection)  occurs at the energies $E_{\rm tr}$ and boundary condition parameters $W_{\rm tr}$ that simultaneously satisfy  the equations  $G_1(E_{\rm tr}) = G_2(E_{\rm tr})$ and $ \Re \, \Sigma(E_{\rm tr})= {W_{\rm tr}}$. The graphs of functions $G_1(E)$ and $G_2(E)$ intersecting at $E=E_{\rm tr}$ are shown in Fig.~\ref{fig:G2}(a). The corresponding value of $W_{\rm tr}$ can be seen in Fig.~\ref{fig:G2}(b). 
The dependence of $E_{\rm tr}$ and $W_{\rm tr}$ on the dielectric constant $\varepsilon_1$ at the semiconductor/vacuum interface ($\varepsilon_2=1$) is shown  in Fig.~\ref{fig:G2}(c,d). 

Surprisingly, perfect transmission may occur exactly at $E_{\rm tr}=V$.
   For  given values of the parameters $\varepsilon_1, \varepsilon_2, m_1, m_2$, the perfect   transmission at $E_{\rm tr}=V$ takes place for $V=V_{\rm tr}$ given by the following formula
\be \label{Vtr}
V_{\rm tr}=\frac{4\pi^2 \mathcal{R}_1}{\ln^2(1+\varepsilon_2/\varepsilon_1)} \, ,  \quad \mathcal{R}_1 = \frac{m_1q_1^2}{2\hbar^2} = \frac{\hbar^2}{2 m_1 \lambda_1^2}.
\ee
For comparison, for $\varepsilon_1=\varepsilon_2$ and $q_{1,2}=0$,  the transmission coefficient  can be written as
\begin{equation}
    T_0(W,E) = \frac{2\hbar^2 \sqrt{E(E-V)}}{\sqrt{m_1m_2}|\Sigma_0(E)-W|^2} \, .
\end{equation}
 The dependence of $T_0(0,E)$ on electron masses $m_1\ne m_2$ and on the barrier height $V$ was studied in Refs.~\cite{Alperovich2021,Alperovich2023}. We note that $T_0(W,V)=0$ while $T_{\rm tr}(W,V)$ is non-vanishing even for small values of $q_1,q_2 \ne 0$.

For $W=0$, several values of $\varepsilon_1$, and $V=1$\, eV  and $V=0.2$\, eV, the dependence of the transmission coefficient on the energy $E>V$  is shown in Fig.~\ref{fig:T}(a) and Fig.~\ref{fig:T}(c), respectively. One can see  that for  $\varepsilon_1=1$  we have $T_0(0,V) =0$  while  $T_{\rm tr}(0,V) \ne 0$ even  for the values of $\varepsilon_1$  very close to one.  In addition, $T_{\rm tr}(0,E)$ changes non-monotonically with $\varepsilon_1$ due to the existence of  perfect transmission points.  Indeed, for  $\varepsilon_1$  such that  the condition $V \approx V_{\rm tr}$ is verified  (see Eq.~\eqref{Vtr}), the value of $T_{\rm tr} \approx 1$ is  nearly constant in the  range of  energies $V<E<V+k_{\rm B}T$  which is painted over by the green color in Fig.~\ref{fig:T}(a) and Fig.~\ref{fig:T}(c).   The dependence  of $  V_{\rm tr}$  on $E-V$  is shown by red and green lines  in Fig.~\ref{fig:T}(a) and Fig.~\ref{fig:T}(c) at $T=300$~K. For other values of $\varepsilon_1$, $T_{\rm tr}$ is usually growing with $E$ in this  energy range.

To analyze the impact of the dielectric confinement on the electron  transmission, we plot the  averaged transmission efficiency, $T^{\rm eff}_{\rm tr}$, representing the transmission coefficient averaged with the Boltzmann distribution of the electron energies $E>V$ (see Eq.~\eqref{eq:Tv} in the next Section) as a function of the surface parameter $W$, several values of $\varepsilon_1$, and $V=1$\, eV  and $V=0.2$\, eV, in  Figures~\ref{fig:T}(b)  and \ref{fig:T}(d).  They show  rapid decrease of  $T^{\rm eff}_{\rm tr}$ with increase of $|W|$.  One can also see   that $T^{\rm eff}_{\rm tr}$ reaches its maximum value   at some value of the surface parameter $W_{\rm tr} \ne 0$. Filled areas in  Figs.~\ref{fig:T}(b)  and \ref{fig:T}(d) show  $T_{\rm tr}$ in the energy range $V<E<V+k_{\rm B}T$. The distribution of the transmission  coefficient,  however, is not observable  for the values of $\varepsilon_1$ close to  the ``optimal'', due to nearly constant value of  $T_{\rm tr} \approx T^{\rm eff}_{\rm tr}$ in this energy range.  

The zero reflection condition, $R_0(W,E_{\rm tr})=1-T_0(W,E_{\rm tr})=0$, could also take place when $k_1(E_{\rm tr})/m_1=k_2(E_{\rm tr})/m_2$ and $W=0$, which means $E_{\rm tr}/m_1=(E_{\rm tr}-V)/m_2$.  One can see that this equality cannot be realized for $m_1<m_2$ and $V>0$. Thus, the appearance   of  zero reflection and perfect transmission at some  energies $E_{\rm tr} \ge V$ as well as the nonzero transmission at $E=V$ are new properties brought by reflection from the image potential $U_{\rm self}$.

\subsection{ Reflection and tunneling under the barrier for $V > E >0$}

\begin{figure}
    \centering
    \includegraphics[scale=1]{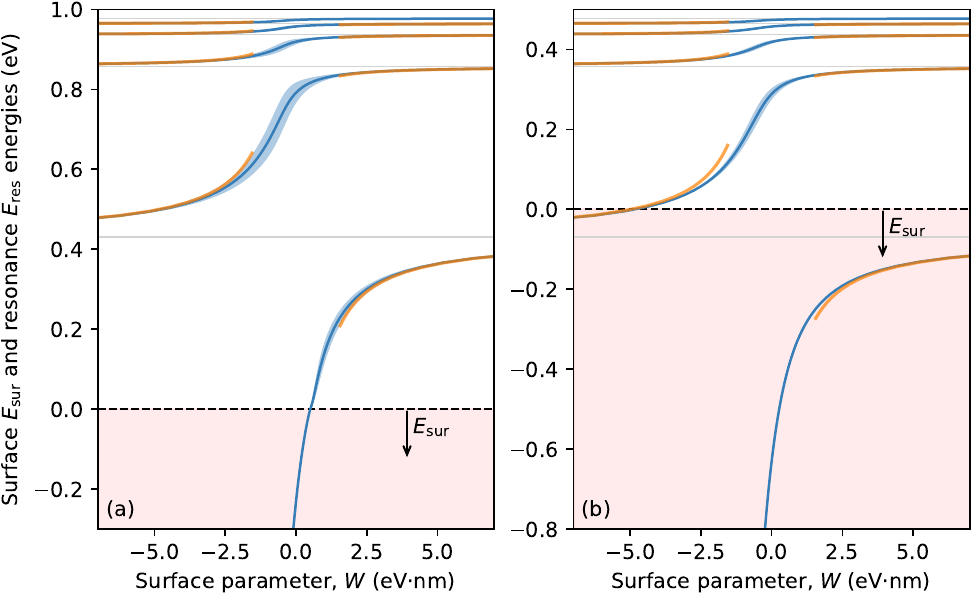}
    \caption{
    Dependence of resonance energies on the surface parameter $W$  calculated  for $V = 1$~eV  (a) and  for $V = 0.5$~eV (b). Imaginary part of $E_{\rm res}$ is shown by the thickness of the blue line. Horizontal lines show the asymptotic resonance energies $E_n$.  The energies  $E_{\rm sur}<0$ below the dashed line correspond to the surface localized states. Orange lines show the resonance energies calculated using asymptotic expression of Eq.~\eqref{eq:E_res_as}  for $E_n^{\rm res}>0$ and of  Eq.~\eqref{Esur_as} for $E_n^{\rm sur}<0$.}
    \label{fig:E_res}
\end{figure}

Next, we consider the energy range $V>E>0$. In this range, the wave function $\psi_2(z)$ decays  for  $z\to +\infty$, and the coefficient $C_2$ vanishes. 
Then, the reflection coefficient is equal to one ($R=1$), and the transmission coefficient  vanishes ($T_{\rm tr}=0$). 

The ratio $r({W}, E) $ of the amplitudes of reflected and falling waves can be written as:
\begin{equation}
   r({W}, E) = \frac{B_1}{ A_1} = \frac{\Gamma(-i\beta_1)}{\Gamma(i\beta_1)}\left(1 - \frac{2iG_1(E)}{\Sigma(E) - {W}}\right).
\end{equation}
 For real values of the energy in the range $V>E>0$, $\Im \Sigma(E) =G_1(E)$, and $r(W,E)$ can be rewritten as 
\begin{equation}
  r({W}, E) =  \frac{\Gamma(-i\beta_1)}{\Gamma(i\beta_1)}\frac{\Sigma^*(E) - {W}}{\Sigma(E) - {W}} = e^{i\phi}
\end{equation}
with the reflection phase $\phi$ given by 
\begin{equation}
     \phi = -2 \arg \Gamma(i\beta_1) - 2\arctan \left(\frac{G_1(E)}{\Re \Sigma(E) - {W}}\right).
\end{equation}

Complex solutions of the equation $\Sigma(E) = {W}$ denoted by $E_{\rm res}$ correspond to resonances of the system. 
The dependence of ${\rm Re} \, \Sigma(E)$  on energy $E$ is shown in Fig.~\ref{fig:Sigma}, and the dependence of ${\rm Re} \, E_{\rm res}$ on the parameter $W$ is shown in  Figs.~\ref{fig:E_res}(a) and  \ref{fig:E_res}(b) for $V=1$~eV and $V=0.5$~eV, respectively.  The imaginary part  ${\rm Im} \, E_{\rm res}$ is negative, and the value of $|{\rm Im} \, E_{\rm res}|$ gives the width of the resonance shown by the thickness of lines in Figs.~\ref{fig:E_res}. Complex roots of the equation $\Sigma(E) = {W}$ are found numerically using the Newton-Raphson method \cite{Press2007}.

\begin{figure}[h!]
    \centering
    \includegraphics[scale=1]{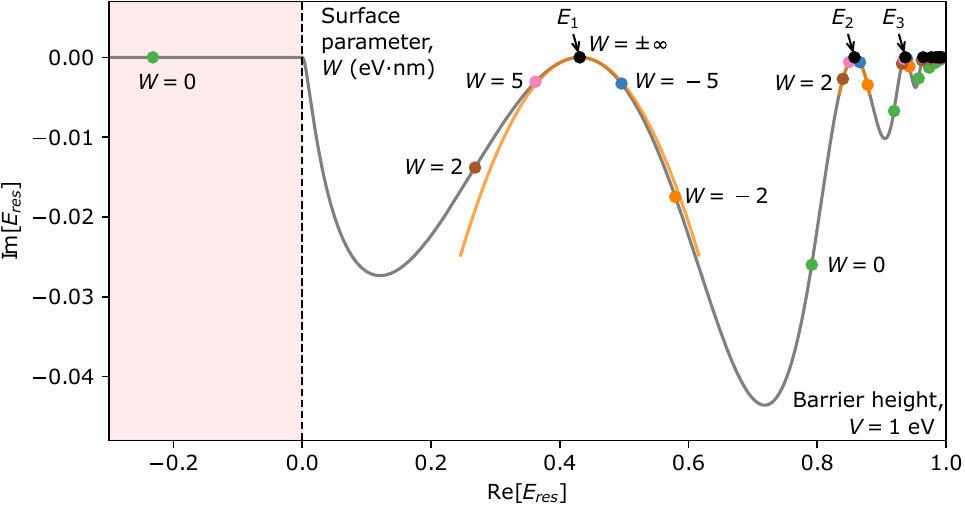}
    \caption{
    Resonance energies on the complex plane for different values of the surface parameter $W$, and for $V = 1$~eV. Orange lines represent the asymptotic formula (\ref{eq:E_res_as}).}
    \label{fig:E_res_compl}
\end{figure}
\begin{figure}[h!]
    \centering
    \includegraphics[scale=1]{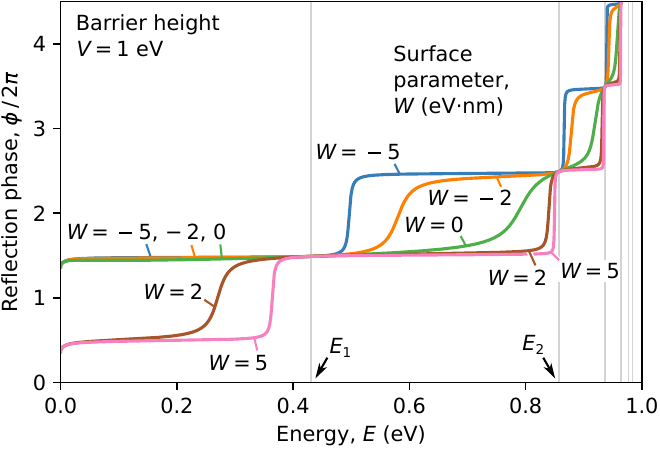}
    \caption{    Dependence of the angle $\phi$ on the energy $E$ calculated for $V = 1$ eV and for different values of the surface parameter $W$.  Vertical gray lines show poles $E = E_n$. }
    \label{fig:phi}
\end{figure}

For $W \to \pm \infty$, one can find the asymptotic expressions for the resonance energies $E_n^{\rm res}$ which are close to the Dirichlet energies $E_n$. 
In more detail, since $\sigma(\alpha_2)$ has poles at $E=E_n$ (see Eq.~\eqref{eq:En}),
so does the function $\Sigma(E)$. Near these poles, and for large values of $|{W}|$,
the function $\Sigma(E)$ can be approximated as follows:
\begin{equation}
    \Sigma(E) \simeq   \frac{2 q_2 \mathcal{R}_2}{n^3 (E - E_n)} + Q(E_n) +i G_1(E_n), \quad E \to E_n,
    \label{eq:B_poles_E}
\end{equation}
where $Q(E_n)$ is the real non-divergent part of $\Sigma(E)$
\begin{align}
    Q(E_n) &= - q_1 \bigl(\ln(2\varkappa_1 \lambda_1) + \sigma(\alpha_1)\bigr) + q_2\Bigl(\ln(2/n) + \sigma(n) - \frac{3}{2n}\Bigr), \quad E < 0, \\
    Q(E_n) &= -q_1 \bigl(\ln(2k_1 \lambda_1) + \Re\sigma(-i\beta_1)\bigr) + q_2\Bigl(\ln(2/n) + \sigma(n) - \frac{3}{2n}\Bigr), \quad 0 < E < V.
\end{align}
Assuming that the resonance energy $E_n^{\rm res}$ is close to $E_n$, one can approximate them as
\begin{equation} 
    E_n^{\rm res} \simeq E_n + \frac{2q_2 \mathcal{R}_2}{n^3 (W - Q(E_n) - i G_1(E_n))} \simeq E_n + \frac{2q_2 \mathcal{R}_2}{n^3  (W - Q(E_n))} + i \frac{2q_2 \mathcal{R}_2 G_1(E_n)}{n^3  (W - Q(E_n))^2}.  \label{eq:E_res_as}
\end{equation}
One can see from Eq.~\eqref{eq:E_res_as} that for $|W - Q(E_n)| \to \infty$, the resonance energy $E_n^{\rm res}$ tends to $E_n$ along a parabola (see Fig.~\ref{fig:E_res_compl}). This is a characteristic behavior of resonances tending to points of the discrete spectrum. Note again that ${\rm Im} \, E_n^{\rm res} < 0 $, and that the quantity $|{\rm Im} \, E_n^{\rm res}|= |2q_2 \mathcal{R}_2 G_1/(n^3 (W - Q(E_n))^2)|$ is the width of the $n$'th resonance. 

For several values of $W$, the dependence of the reflection phase angle $\phi$ on the electron energy is shown in Fig.~\ref{fig:phi}. Note that the phase $\phi$ experiences jumps of $2\pi$ near resonances.

\subsection{Surface states  
below the semiconductor band edge}

Surface states  can be created by the attractive  mirror force potential   below the semiconductor band edge $E=0$. 
The condition for the wave function to decay for $z \to \pm \infty$ on both sides of the interface   yields $C_1=C_2=0$. 
We start with the analysis of Dirichlet boundary conditions $\psi_1(0)=\psi_2(0)$ which correspond to the limit $|{W- Q(E_n)}| \to \infty$. In that case, the isolated eigenvalues arise for $z>0$, and they correspond to the poles of the function $\Gamma(\alpha_2)$ at the points $\alpha_2=-n$, and to the Dirichlet 
 energies $E_n$ given by Eq.~\eqref{eq:En}.

For finite values of ${W}$, the discrete spectrum is again given by  solutions of the equation $\Sigma(E) = {W}$. The behavior of the function $\Sigma(E)$ for $E<0$ is shown in Fig.~\ref{fig:Sigma}. In the energy range $E<0$, solutions of the equation $\Sigma(E) = {W}$ denoted by $E_{\rm sur}$ are shown in Fig.~\ref{fig:E_res}. Real roots of the equation $\Sigma(E) = {W}$ for $E<0$ are found numerically using the Newton-Raphson method \cite{Press2007}.

In the case of $\mathcal{R}_2 >V$, some of the Dirichlet energies $E_n$ might be negative (Fig.~\ref{fig:E_res}(b)). They correspond to the poles of the function $\sigma(\alpha_2)$ for negative integer values of $\alpha_2$. At the same time, the values of $\alpha_1$ are positive and the function $\sigma(\alpha_1)$ does not have poles. Using the same asymptotic analysis as in the previous section, we obtain the following asymptotics valid for large values of $|W- Q(E_n)|$:
\begin{equation} \label{Esur_as}
    E_n^{\rm sur} \approx  E_n + \frac{2\mathcal{R}_2 }{n^3} \frac{q_2}{W- Q(E_n)}.
\end{equation}
This formula can be compared to Eq.~(\ref{eq:E_res_as}), where $G_1(E)=0$ for $E<0$.

\section{Applications: photoemission and surface quantum well}
\label{Sec:5}

In this Section, we consider two applications of our approach 
to  heterostructures of the type shown in Fig.~\ref{fig:potential}.
The first concerns the photoemission of electrons from the semiconductor to vacuum,  and the second   is the study of quantum states confined  in a quantum well located  near the surface of a semiconductor. While the discontinuity of the self-interaction potential affects multiple phenomena in optical and electrical properties of nanostructures, we selected  these two effects because they can be described within the one-dimensional model developed above.

\subsection{Electron photoemission}

 \begin{figure}[h!]
    \centering
    \includegraphics[scale=1]{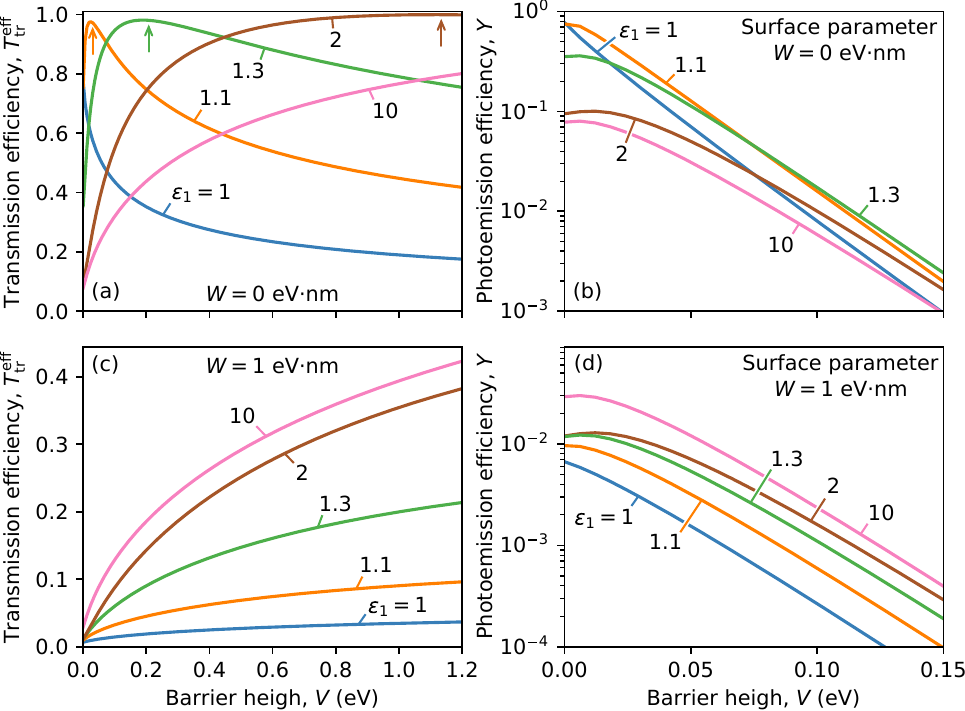}
    \caption{
    Dependencies of the transmission efficiency $T^{\rm eff}_{\rm tr}$ (a,\,c) and  photo-emission efficiency $Y$ (b,\,d) on the barrier height $V$ calculated for the surface parameter $W=0$ (a,\,b)  and  for $W=1$~eV$\cdot$nm (c,\,d)  with different values of the dielectric constant $\varepsilon_1$, and  for $\varepsilon_2=1$. Arrows in (a) show the values of the ``optimal'' barrier height $V_{\rm tr}$ from Eq.~\eqref{Vtr}.}
    \label{fig:Teff} 
\end{figure}

First, we address the influence of the dielectric discontinuity on photoemission.   In our model,  we neglect a  semiconductor doping which leads to band bending near the semiconductor/vacuum interface  and to acceleration of  electrons falling on the surface (see Fig.~1 in \cite{Alperovich2023}).   Consequently, only electrons with kinetic energy larger than band offset $V$ contribute to the photoemission. The quantum efficiency $Y$ of electron  photoemission  is defined in \cite{Alperovich2021,Alperovich2023} as the  ratio of the emitted electron flux, taking into account the transmission coefficient $T$, to the total flux of electrons incident on the surface from the semiconductor volume. Considering only the normal incidence of electrons on the semiconductor/vacuum interface, we obtain
\be
Y(W,V)={\int_0^\infty T_{\rm tr}(W,E) k_1(E) e^{-E/k_{\rm B}T}  \dd  k_1 \over \int_0^\infty k_1(E) e^{-E/k_{\rm B}T}  \dd  k_1}.
 \ee
Here we assume that electrons are thermalized, and that their thermal-population is described by the Boltzmann distribution  $f_B(E)=\exp(-E/k_{\rm B}T)$, where $E= \hbar^2k_1^2/2m_1$ is the electron energy, and $k_B$ is the Boltzmann constant.    One can see that in our one-dimensional model
the quantum efficiency of the photoemission can be expressed  as follows:
\begin{equation}
Y(W,V) =  \exp{-\frac{V}{k_{\rm B}T}} T^{\rm eff}_{\rm tr}(W),  \hskip 0.3cm
T^{\rm eff}_{\rm tr}(W)= \frac{\int_V^\infty   T_{\rm tr}(W,E)e^{-E/k_{\rm B}T} \dd E}{ \int_V^\infty  e^{-E/k_{\rm B}T} \dd E}. \label{eq:Tv}
 \end{equation}
 Here $T_{\rm tr}^{\rm eff}(W)$ is the transmission efficiency. The dependence of $T^{\rm eff}_{\rm tr}(W)$ on the surface parameter $W$ for  $\varepsilon_2=1$, various values of $\varepsilon_1$, and the  barrier heights $V=1$~eV and $V=0.2$~eV  is shown  in Fig.~\ref{fig:T}(b)  and  Fig.~\ref{fig:T}(d), respectively.  One can see that $T^{\rm eff}_{\rm tr}(W)$, and consequently $Y(V, W)$, significantly reduces with increase of $|W|$. The maximum of $T^{\rm eff}_{\rm tr}$ for a given value of $\varepsilon_1$ is reached at the non-zero (but usually small) value $W=W_{\rm tr}$. Interestingly, for a  fixed value of the surface parameter $W$ and for $V>0.18$~eV, the value $T^{\rm eff}_{\rm tr}$ taking into account the self-interaction potential $U_{\rm self}$ ($\varepsilon_1>\varepsilon_2$) is larger than  that for $q_1,q_2 =0$.

Figure \ref{fig:Teff} shows the dependencies of the transmission efficiency $T^{\rm eff}_{\rm tr}(W)$ (a,\,c) and photoemission efficiency $Y(W,V)$ (b,\,d) on the barrier height $V$ calculated for $W=0$ (a,\,b) and $W=1$~eV$\cdot$nm (c,\,d) with $\varepsilon_2=1$ and various values of $\varepsilon_1$.  Unexpected  non-monotonic  dependence of $T^{\rm eff}_{\rm tr}(W)$ on  $V$  for $W=0$   can be observed close to the optimal conditions.  Figure \ref{fig:Teff}  shows also  that for  $W=1$~eV$\cdot$nm (that is, far from the resonant condition)  $T^{\rm eff}_{\rm tr}(W)$  unexpectedly increases with increase of $V$. Both these phenomena result from the perfect transmission conditions   predicted  by our theory.  For $Y(V)$, the main tendency is the decrease of $Y$ with  the  increase of $V$ because  of the exponential factor in the Boltzmann distribution. However, one can see that taking into account the potential $U_{\rm self}$ gives rise to a non-monotonic dependence of $Y$ on $V$ for small values of $V$. Depending on the barrier height $V$ and the value of the surface parameter $W$, an additional scattering on the mirror force potential $U_{\rm self}$ may  increase or decrease the quantum efficiency of the photoemission.

\subsection{Discrete spectrum of electrons in a surface quantum well}
 
 \begin{figure}[h!]
    \centering
    \vspace{-0.5cm}
    \includegraphics[scale=1]{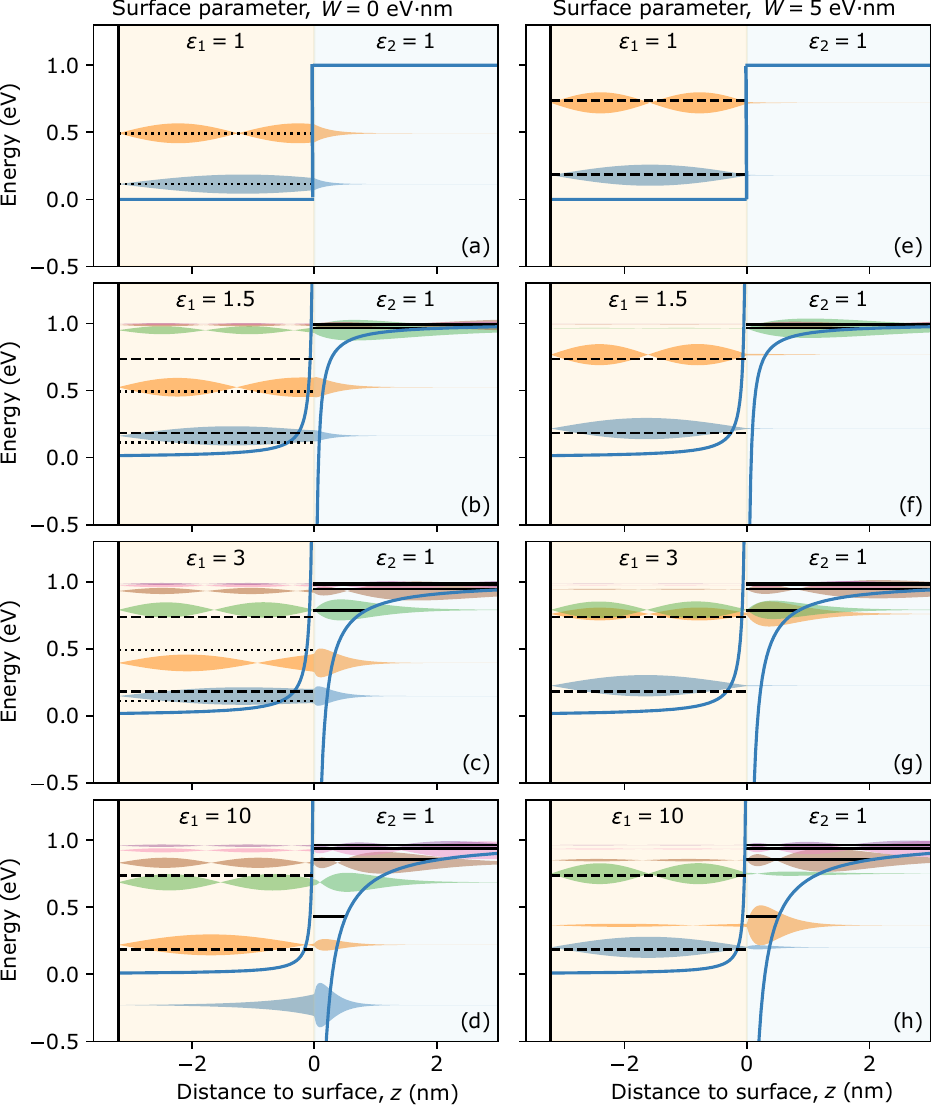}
    \caption{Electron states in the asymmetric quantum well calculated  for  various dielectric constants $\varepsilon_1$.  Solid blue lines show the electron potential $U(z)$. Horizontal solid lines show the energy levels  $E_n$ localized in the  mirror force attractive potential; dashed lines --- levels $E_{n}^\infty$ in the QW with infinite barriers;  dotted lines --- levels $E_{n}^0$ in the asymmetric QW for $q_1,q_2=0$. Filled areas show the electron wave functions at the energy levels. 
    \looseness=-1}
    \label{fig:QW}
\end{figure}

Next, we consider a quantum well (QW) between an infinite potential barrier at $z=-a$ and  the discontinuity of the dielectric constant at $z=0$. The boundary condition at $z=-a$ is given by $\psi(-a) = 0$ while GBC is used for  $z=0$. 
We focus our attention on the discrete spectrum of energy levels with $E<V$. 

In the absence of dielectric  discontinuity (the case of $q_1,q_2=0$), the discrete spectrum $E_n^{W}$ is given by  solutions of the equation:
\be \label{EnuW}
\left( 1+ \frac{2W}{\hbar^2}\frac{m_2}{\varkappa_2}  \right) \tan(k_1a) + \frac{k_1}{\varkappa_2} \frac{m_2}{m_1} = 0 \, .
\ee
For $|W| \to \infty$, we obtain $E_n^{W} \to E_n^\infty$, where
$$
E_n^\infty  = \frac{\hbar^2 \pi^2 n^2}{2m_1 a^2} \, ,  \quad n=1,2,3,\ldots
$$
are the energy levels in the symmetric QW with two infinite potential barriers.
In this case, the surface parameter $W$ can be directly interpreted as the power of the $\delta$-function potential at $z=0$ preventing electron tunneling.

 In the absence of dielectric  discontinuity and for $W=0$,  the number $\nu$ of confined levels, $E_n^{0}$ with $n=1,2,\ldots,\nu$,  localized in the asymmetric quantum well with a finite value of the band offset potential $V$,  can be estimated by the condition  
\be
a \sqrt{2m_1 V} \ge \hbar \pi  (\nu -1/2),
\ee
where $a$ is the quantum well width.  Selecting a quantum well containing two quantum  confined levels with energies $E_1^0$ and $E_2^0$ in the case of  $q_1,q_2, W=0$, and using the following material parameters: $m_1=0.2m_0$,  $V=1$~eV  and  $a=3.2$~nm,  we will investigate the effect of dielectric confinement and of the surface parameter $W$ on the electron level structure.

In Figure \ref{fig:QW}, we show the electron energy spectrum and the profile of wave functions for such a well taking into  account  the mirror potential $U_{\rm self}$ for various values of $\varepsilon_1=1$ (a,\,e), 1.5 (b,\,f), 3 (c,\,g) and 10 (d,\,h). The energy states $E_n^{\rm QW}$ are found by solving the equation $\psi_1(-a) = 0$ assuming that $C_2 = 0$ to ensure that $\psi_2(z)$ decays for $z\to +\infty$. The equation $\psi_1(-a) = 0$ is solved numerically using Brent's method \cite{Press2007} in the energy range $-1\,\text{eV} < E < V$, which is subdivided into 100 uneven intervals that narrow near $E = V$. This ensures that the lowest 10 energy levels are certainly found.

The values of the surface parameters are $W=0$~eV$\cdot$nm in the panels (a,\,b,\,c,\,d) and $W=5$~eV$\cdot$nm in the panels (e,\,f,\,g,\,h). Horizontal solid lines show the energies $E_n = V - \mathcal{R}_2/n^2$ (see Eq.~\eqref{eq:En}) given by  the poles of  the function $\Sigma(E)$ and being the levels created by mirror force potential attracting electron to the surface in the case of Dirichlet boundary conditions. Horizontal  dashed lines show  the energy levels $E_{n}^\infty$ in the symmetric QW with infinite barriers at $z=-a$ and $z=0$ for $q_1,q_2=0$. Horizontal  dotted lines show the energy levels $E_{n}^0$ in the asymmetric QW  with finite potential barrier $V=1$~eV at $z=0$ for $q_1,q_2,W=0$. Filled areas show the profiles of electron wave functions corresponding to the localized states.

 Comparing the panels (a) and (e), one can observe that for $q_1,q_2=0$ the boundary conditions with $W=5$~eV$\cdot$nm  shift two electron energy levels in the QW from $E_{n}^0$  to $E_{n}^\infty$, and that the electron tunneling under the barrier is very weak. Even a small dielectric confinement (panels (b) and (f)) modifies the energy structure:  resonant levels near the barrier appear with electron wave functions mostly localized in the dielectric area. With increase of the dielectric confinement, the levels with electrons localized both in the QW and in the dielectric appear including a surface localized state with $E<0$ (see panel (d)).

\begin{figure}[tb!]
    \centering
    \includegraphics[scale=1]{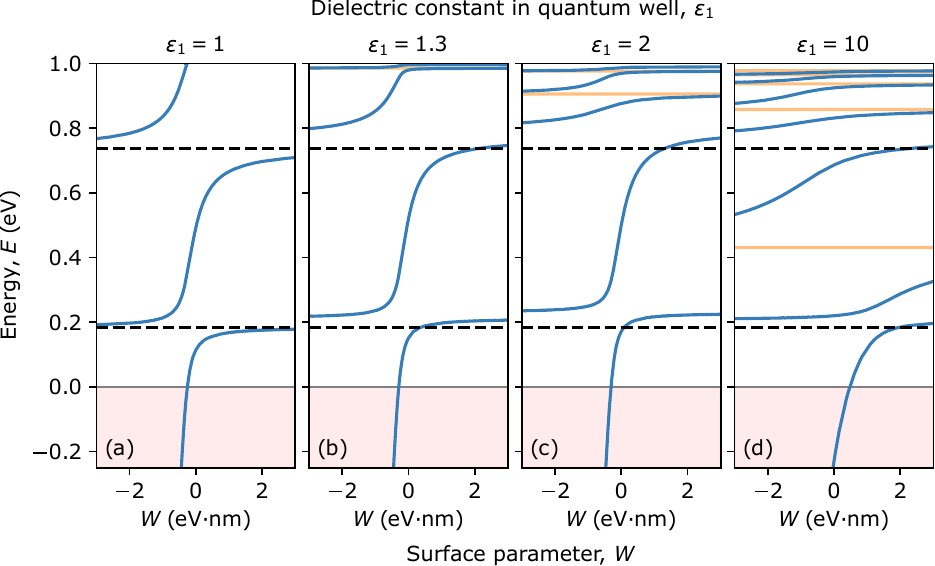}
    \caption{ Dependence of the electron energy levels in the asymmetric quantum well on the surface parameter $W$  calculated for  dielectric $\varepsilon_2=1$  and dielectric constants $\varepsilon_1=1$, 1.3, 2, 10 in panel (a) to (d), respectively.  The levels are shown by a blue lines.
    Horizontal  dashed lines show the energy levels $E_{n=1,2}^\infty$ in the symmetric QW with infinite barriers for $q_1,q_2=0$,   orange solid lines show the pole energies $E_n$ for levels localized in the attractive potential of mirror force outside of semiconductor.}
       \label{fig:QW_states_W} 
\end{figure}

\begin{figure}[tb!]
    \centering
    \includegraphics[scale=1]{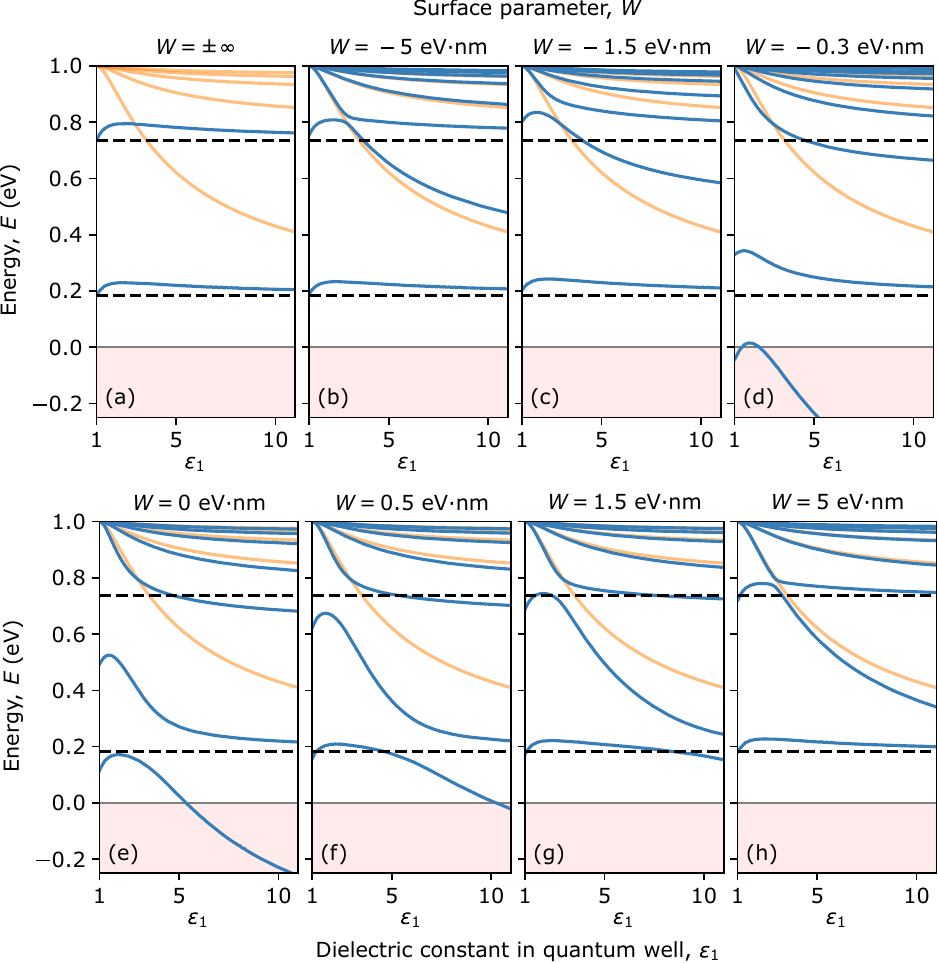}
    \caption{ Dependence of the electron energy levels shown by blues lines  in the asymmetric quantum well as a function  of  the dielectric constant $\varepsilon=1$, calculated for $\varepsilon_2=1$  and  surface parameter $W=\pm\infty$ and $W=-5$, $-1.5$, $-0.3$, $0$, $0.5$, $1.5$, $5$\,eV$\cdot$nm in panels (a)  to (h), respectfully. Orange lines show the pole energies $E_n$ for levels localized in the attractive potential of mirror force outside of semiconductor.  Horizontal  dashed lines show the energy levels $E_{n=1,2}^\infty$ in the symmetric QW with infinite barriers for $q_1,q_2=0$.}
      \label{fig:QW_states} 
\end{figure}

Figure \ref{fig:QW_states_W} shows the dependence  of  electron energy levels (blue lines) on the surface parameter $W$ for various values of $\varepsilon_1$.   Horizontal  dashed lines show energy levels $E_{n}^\infty$ in the symmetric QW, 
orange lines  -- energy levels $E_n = V - \mathcal{R}_2/n^2$. One can see that even without dielectric confinement for $\varepsilon_1=\varepsilon_2=1$ (Fig.~\ref{fig:QW_states_W}(a)) the energy levels $E_{n}^W$ in a QW can be dramatically modified by the presence of the surface short-range potential $W\delta(z)$.  The dielectric confinement with $\varepsilon_1 >\varepsilon_2$ (Fig.~\ref{fig:QW_states_W}(b,\,c,\,d)) has an additional effect  of mixing  the $E_{n}^W$  QW states with the $E_n$ states in the attractive potential $U_{\rm self}(z)$. 

Figure \ref{fig:QW_states} shows the dependence of  electron energy levels on the value of $\varepsilon_1$  for various values of the surface parameter $W$. The energy levels in the quantum well are shown by blue lines, while the orange lines show the energies $E_n = V - \mathcal{R}_2/n^2$. Horizontal  dashed  lines show energy levels $E_{n}^\infty$. Blue lines are plotted as a contour plot $\psi(-a) = 0$ on the plane $(\varepsilon_1, E)$, employing the adaptive sampling method to enhance numerical performance. 

As discussed above, the case of $W \to \pm \infty$  (Fig.~\ref{fig:QW_states}(a)) corresponds to the Dirichlet boundary conditions $\psi_1(0)=\psi_2(0)=0$. Correspondingly, the two electron states $E_{n=1,2}^{\rm QW}$ are fully localized in the QW while the $E_n$ states are localized in the attractive potential $U_{\rm self}(z)$  in the dielectric. The dielectric confinement with $\varepsilon_1 \ne \varepsilon_2$ weakly affects the energies $E_{n=1,2}^{\rm QW} \approx E_{n=1,2}^\infty$ by shifting them up. However, for the finite values of $W$ allowing the electron tunneling under the barrier, the resulting energy states are  superposition states of the electron in the quantum well and in the attractive potential $U_{\rm self}(z)$  in the dielectric.  It is worth noting that in this case the dramatic modification of the energy spectrum takes place already for $\varepsilon_1<2$,  and that the potential $U_{\rm self}(z)$ cannot be considered as a perturbation. 

\subsection{Discrete spectrum of electrons in a quantum well with high-k dielectric barrier}

\begin{figure}[tb!]
    \centering
    \includegraphics[scale=1]{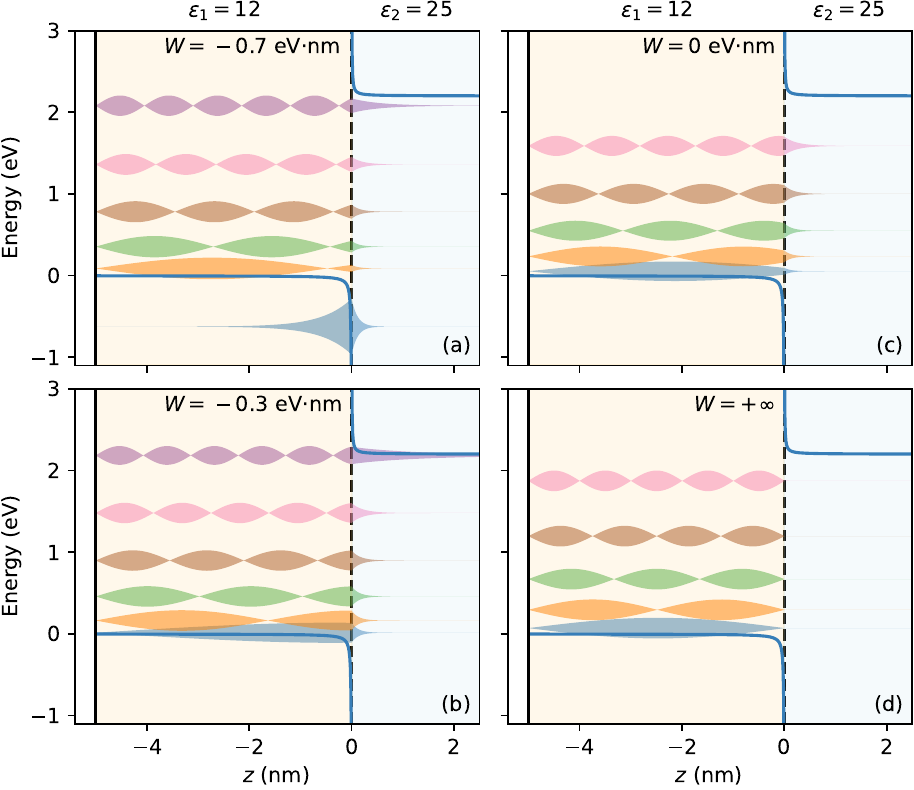}
    \caption{Electron levels in a 5nm thick asymmetric QW with a high-k dielectric barrier calculated for $W=-0.7$, $-0.3$, $0.0$, and $+\infty$ eV$\cdot$nm are shown in the panels (a) to (f), correspondingly. Solid blue lines show the electron potential $U(z)$. Vertical dashed lines indicate the interface between the QW and the high-k dielectric. Filled areas show the wave functions of  the corresponding   energy  levels.} 
    \label{fig:high-k-QW}
\end{figure}

In this Section, we consider an asymmetric quantum well having an infinite potential barrier without dielectric discontinuity on the left and a finite barrier ($V=2.2$ eV) between the semiconductor and the high-k dielectric on the right $z=0$. In this case,  $\varepsilon_1 < \varepsilon_2$. Such  Si/high-k interfaces, are important in modern field-effect transistors \cite{GAAFET}.  The dielectric contrast between a semiconductor ($\varepsilon_1=12$) and a high-k dielectric ($\varepsilon_1=25$) creates a potential shown in Fig.~\ref{fig:high-k-QW}. The values of the electron energy barrier and of the dielectric contrast used in our numerical simulations are typical for a Si/HfO$_2$ interface \cite{Tan2009, Robertson2004}.

 Figure \ref{fig:high-k-QW}  shows the exact  electron energy levels calculated in a 5 nm thick surface quantum well for  different values of the surface parameter $W$. The largest value of $W=+\infty$ corresponds to the case of an impenetrable barrier with the Dirichlet boundary condition $\psi(z=0)=0$.    One can observe that depending on the value of $W$  the number of  quantum size levels localized in the QW changes, and that surface-localized states may appear for $W<0$. The value of $W$ also affects penetration of electron wave functions into the high-k dielectric layer, and it may potentially can increase  its capacitance.
 
\section{Discussion and Conclusion} \label{Sec:discussion}

In this paper, we have presented a non-perturbative macroscopic theory for flat interfaces with a dielectric discontinuity.
It gives rise to a  self-consistent quantitative description of  the  leakage  of the carrier wave function in the surrounding matrix, and such a description is crucial for the study of transport in nanocrystal arrays and solids. This theory also predicts several novel phenomena.  The first one is the existence of perfect transmission conditions above the barrier, and of resonance  levels with finite life time below the barrier that could potentially enhance  photo-effect in some heterostructures.  For surface quantum wells, it predicts strong coupling between internal confined levels and surface levels created by the attractive mirror potential for some interfaces.  As a consequence, conductivity of an array of such structures should be enhanced. 
 Furthermore, in addition to engineering of the potential barrier, engineering of the dielectric discontinuity provides a new tool controlling transitions between bound states localized in quantum wells and resonances in the continuous spectrum \cite{DasSarma2024}. 

Our non-perturbative theory  describes macroscopic  properties of the interface layer  using a single phenomenological parameter $W$. 
The value of  $W$ can be found  from first principle calculations  of the electronic properties of the interface, or from fitting experimental characteristics of the interface with our theory.

\begin{figure}[h!]
    \centering
    \includegraphics[scale=1]{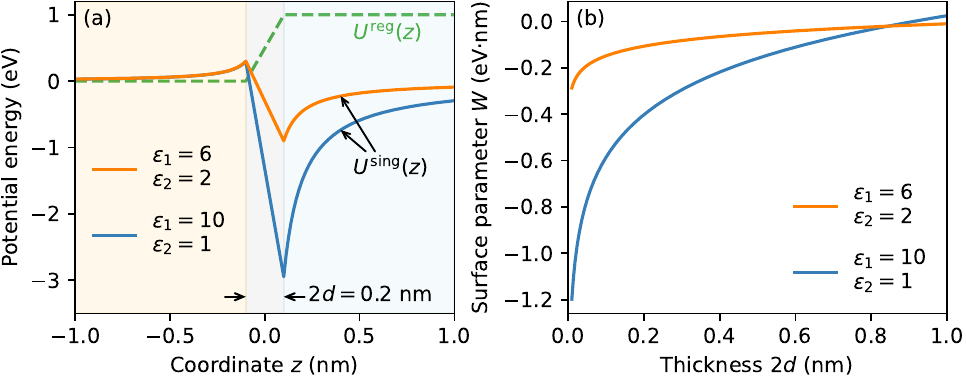}
    \caption{  (a) Singular self-interacting potential, $U_{\rm sing}(z)$, created by mirror charge for parameters from Figs.~\ref{fig:potential}(a) (blue curve) and \ref{fig:potential}(b) (orange curve)  with the linear interpolation in the interface layer of the width $2d$. Green dashed line shows the regular potential energy $U^{\rm reg}(z)$ due to the barrier height $V=1$ eV.   
    (b) The dependencies of the corresponding surface parameter $W$ on the interfacial thickness $2d$.}
      \label{fig:linear} 
\end{figure}

The parameter $W$ can also be estimated  using Eq.~\ref{eq:W} if  we assume  some microscopic  behavior of the singular interface potential $U^{\rm sing}$. For example, for a linear interpolation of the self-interaction potential in the interface layer extending from $-d$ to $d$ (see Fig.~\ref{fig:linear}(a)), the surface parameter $W$ is given by
\begin{equation}
    W =  q_2 (1+ \ln(\lambda_2/d))-q_1(1+ \ln(\lambda_1/d)).
   \label{eq:63}
    \end{equation}
Such a linear interpolation, together with a linear interpolation of the step-like potential $U^{\rm reg}$,  was assumed in Refs.~\cite{MuljarovSPIE1993,MuljarovPRB1995}, where  the  thickness $2d$ of the interface layer was considered as an adjusting parameter. The dependence of $W$ on $2d$ is calculated using Eq.~\eqref{eq:63}
is shown in Fig.~\ref{fig:linear}(b).
 This dependence suggests a  range of  physically justified  values of $W$  for a given contrast of the dielectric constants. Note that in the linear interpolation model  $2d$ cannot be chosen smaller than the atomic size, 
and $d$ should not  
be larger  than  the typical depth $\lambda_V = \sqrt{\hbar^2/2m_2V}$ of the electron tunneling under the barrier $V$ created by the band offset between the two solids. 
In the case of the semiconductor/high-k interface shown in Fig.~\ref{fig:high-k-QW}, the estimation of the surface parameter $W$ with the help of Eq.~\eqref{eq:63} for the interface thickness $2d$ between 0.1 and 1 nm gives relatively low values of $W$ ranging from $-0.033$ eV$\cdot$nm to $-0.046$ eV$\cdot$nm, respectively. At the same time, the surface parameter $W$ may depend on the microscopic structure of the surface, which was not taken into account in Eq.~\eqref{eq:63}.


We can also apply our theory  to determine the value of $W$ for the liquid helium /vacuum  interface.  Due to its large size and ideal periodic structure, this  interface is most heavily studied  experimentally. 
The self-interaction potential of an electron near the liquid helium surface leads  to a series of localized  electron levels, which are  shown schematically  in Fig 1(a).  The first experimental study of these  levels  
was reported in Ref.~\cite{Grimes1976}.  In more detail, they  performed spectroscopic measurements of transition energies between the excited states with $n=2,3$ and the ground state with $n=1$, and they obtained the following results:
\begin{equation}
    E^{\rm exp}_{1 \rightarrow 2} = 125.9 \pm  0.2\ {\rm GHz} = 0.521\ {\rm meV},
    \hskip 0.3cm
    E^{\rm exp}_{1 \rightarrow 3} = 148.6 \pm 0.3\ {\rm GHz} = 0.615\ {\rm meV}.
\end{equation}
In the framework of this paper, these transitions correspond to $E_{1 \rightarrow 2} = E_{n=2}^{\rm sur} -  E_{n=1}^{\rm sur}$ and $E_{1 \rightarrow 3} = E_{n=3}^{\rm sur} -  E_{n=1}^{\rm sur}$, respectively. As it was noticed in Ref.~\cite{Grimes1976}, the surface energies $E_n$ obtained using Dirichlet boundary conditions with $\varepsilon_1 = 1.05723$ and  $\varepsilon_2 = 1$ give energy separations which are about 5\%  smaller than  the experimental values.
Several  theoretical works fitted  experimental data by considering a modified 
image charge potential, or microscopic details of the interface \cite{Grimes1976,Frank,Hipolito1978,Cheng1994}. In our model,  using  $\varepsilon_1 = 1.05723$, $\varepsilon_2 = 1$, $V=-1$~eV  from \cite{Grimes1976}  we fit the experimental transitions as
\begin{equation}
    E^{\rm theor}_{1 \rightarrow 2}= 0.52149\ {\rm meV},
    \hskip 0.3cm
    E^{\rm theor}_{1 \rightarrow 3}=0.61562\ {\rm meV}
\end{equation}
using the value of the surface parameter $W=0.26$ eV$\cdot$nm. The energies $E_{n}^{\rm sur}$ were found numerically as  solutions of the equation $\Sigma(E_{n}^{\rm sur})=W$.  The asymptotic  expression of Eq.~\eqref{Esur_as} also gives a very good description of level splitting with an accuracy of $\pm 0.0005$\ meV.
This excellent agreement with experiment demonstrates that our non-perturbative microscopic theory of interfaces with discontinuous dielectric constant describes properties of these interfaces without  modification of the image charge potential, or knowledge of microscopic details of the interface.

The one-dimensional non-perturbative  theory developed in this paper  is only the first step in a self-consistent  quantitative description of electronic and optical properties of nanostructures with dielectric discontinuities. 
We have focused on the carrier motion in the direction perpendicular to the surface, and
we have neglected the carrier motion along the surface. For this reason, our considerations do not include effects of the dielectric discontinuity on the interface polariton modes which are discussed in Ref.~\cite{Ridley}.  We have also assumed that the non-local character of dielectric permittivity at the scale of individual atoms can be taken into account by the surface parameter $W$.
Furthermore, we have taken into account  the instantaneous part of the dielectric permittivity neglecting  electronic and lattice relaxation processes in the surrounding medium. 

A natural next step is to apply our theory to  nanostructures  with high enough symmetry  allowing for an exact analytic expression for the mirror potential, and for  separation of variables in the corresponding Schrödinger equation.
These  include spherical and  ellipsoidal  nanocrystals, nanorods, nanowires, and nanoplatelets.  In such structures,  the non-perturbative theory is expected to  provide a complete realistic description  of quantum confined and surface localized levels as well as the wave function leakage into the barrier, of the oscillator transition strength of the optical transition, of the electron-hole exchange interaction, {\em etc.}  More ambitiously,  the GBC of the type derived in this paper for interfaces with discontinuity of dielectric  constants  should allow for a qualitative numerical  description of all nanostructures.

\begin{acknowledgments}
We are indebted to L. Parnovski for pointing out the reference \cite{Golovaty2019}, for the detailed explanations of the results of that paper, and for many valuable suggestions. We are grateful to  M. Dykman, J. Galkowski,  N. A. Gippius, A. A. Golovatenko, J. Lyons, M. Swift, S. G. Tikhodeev, and C. White for inspiring discussions. 

Y.M.B. acknowledges support provided by the Ioffe Institute program FFUG-2024-0037. 
Research of A.V.R. and A.A. was supported in part by the grants 208235, 220040, and 236683, and by the National Center for Competence in Research (NCCR) SwissMAP of the Swiss National Science Foundation. A.A. acknowledges support of  the award of the Simons Foundation to the Hamilton Mathematics
Institute of the Trinity College Dublin under the program ``Targeted Grants to Institutes''.
Al.L.E acknowledges support by the Office of Naval Research through the Naval Research Laboratory’s Basic Research Program.
\end{acknowledgments}

\section*{Data Availability Statement}

The data that support the findings of this study are available from the corresponding author upon reasonable request.



\end{document}